\documentclass[10pt,journal,compsoc]{handoff/IEEEtran}
\ifCLASSOPTIONcompsoc
\usepackage[nocompress]{cite}
\else
\usepackage{cite}
\fi
\usepackage{graphicx}
\usepackage{pdfpages}
\usepackage{pax}
\makeatletter
\def\PAX@attrs@GoTo#1{#1}
\makeatother
\graphicspath{{fig/}}
\usepackage{amsmath,amssymb,amsfonts}
\usepackage{amsthm}
\usepackage{algorithm}
\usepackage{algorithmic}
\usepackage{array}
\usepackage{textcomp}
\usepackage{xcolor}
\usepackage{multirow}
\usepackage{rotating}
\usepackage{hyperref}
\usepackage{ragged2e}
\usepackage{float}
\allowdisplaybreaks

\begin{document}
%
\title{Quantum Generative Diffusion Model: A Fully Quantum-Mechanical Model for Generating Quantum State Ensemble}
%
%
%
%

\author{Chuangtao~Chen,
		Qinglin~Zhao,~\IEEEmembership{Senior~Member,~IEEE,}
        MengChu~Zhou,~\IEEEmembership{Fellow,~IEEE,}
        Zhimin~He,
        Zhili~Sun,~\IEEEmembership{Senior~Member,~IEEE,}
        and~Haozhen~Situ
\thanks{Corresponding author: Qinglin Zhao.}
\IEEEcompsocitemizethanks{\IEEEcompsocthanksitem Chuangtao Chen and Qinglin Zhao are with the Faculty of Innovation Engineering, Macau University of Science and Technology, Macao 999078, China. (e-mail: chuangtaochen@gmail.com; qlzhao@must.edu.mo)
\IEEEcompsocthanksitem MengChu Zhou is with the Department of Electrical and Computer Engineering, New Jersey Institute of Technology, Newark, NJ 07102 USA. (e-mail: zhou@njit.edu)
\IEEEcompsocthanksitem Zhimin He is with the School of Electronic and Information Engineering, Foshan University, Foshan 528000, China. (e-mail: zhmihe@gmail.com)
\IEEEcompsocthanksitem Zhili Sun is with the Institute of Communication Systems, University of Surrey, GU2 7XH Guildford, U.K. (e-mail: Z.Sun@surrey.ac.uk)
\IEEEcompsocthanksitem Haozhen Situ is with the College of Mathematics and Informatics, South China Agricultural University, Guangzhou 510642, China. (e-mail: situhaozhen@gmail.com)
}
}

\IEEEtitleabstractindextext{%
\begin{abstract}
		Mixed quantum states are the native description of many physically important quantum systems, making their generation a fundamental task in quantum information processing. However, constructing a diffusion process that generates density operators while keeping every reverse step physically valid remains nontrivial.
		This work introduces Quantum Generative Diffusion Model (QGDM), a fully quantum-mechanical model whose forward and backward processes are grounded in quantum channel theory.
		Through a non-unitary forward process, any target quantum state can be transformed into a completely mixed state.
		A trainable backward process recovers the former from the latter. 
		We introduce partial trace to make the backward process non-unitary, and share parameters across timesteps by incorporating temporal information as an input.
		We present QGDM's resource-efficient version to reduce auxiliary qubits while preserving generative capabilities.
		We theoretically analyze the denoising design, showing it avoids a low-loss shortcut that traps training and cause generation failure.
		Simulations confirm that QGDM outperforms quantum generative adversarial networks on random pure- and mixed-state generation, with better noise robustness than other quantum generative models and a task-specialized approach for practical Gibbs state generation. 
		Hence, QGDM provides a channel-based diffusion framework for learning fixed mixed-state targets, extending quantum generative modeling toward realistic quantum information settings.

\end{abstract}

}

\maketitle

\IEEEdisplaynontitleabstractindextext

%
\IEEEpeerreviewmaketitle

\IEEEraisesectionheading{\section{Introduction}\label{sec:introduction}}

\IEEEPARstart{Q}{uantum} state preparation and generation are fundamental tasks in quantum computation, quantum simulation, and quantum-information processing~\cite{nielsen2010quantum,wilde2013quantum}. Within this broad task, mixed-state generation is particularly important because density operators naturally describe a wide range of physically relevant quantum scenarios. For instance, they describe thermal equilibrium with a heat bath~\cite{nielsen2010quantum,wilde2013quantum}, open-system dynamics under environmental coupling~\cite{zurek2003decoherence}, and noisy outputs produced by near-term quantum devices~\cite{braccia2021enhance,consiglio2024variational}. Unlike ordinary data generation, mixed-state generation must respect the density-operator structure. The generated object must remain positive semidefinite with unit trace, and the transformations used during generation should be physically valid quantum operations.

Quantum generative models~\cite{tian2023recent} provide several routes for representing or sampling quantum and classical distributions within quantum machine learning~\cite{biamonte2017quantum,schuld2019quantum,shi2022parameterized}. Among the most studied frameworks, Quantum Generative Adversarial Networks (QGANs)~\cite{lloyd2018quantum,dallaire2018quantum} use an adversarial generator--discriminator structure to generate quantum states or fit classical data distributions~\cite{zoufal2019quantum,situ2020quantum,huang2021experimental,silver2023mosaiq,tsang2023hybrid,zhou2023hybrid}. Quantum Circuit Born Machines (QCBMs)~\cite{benedetti2019generative,liu2018differentiable} train parameterized quantum circuits to sample classical bit-string distributions, with applications in combinatorial optimization and distribution learning~\cite{coyle2020born,kiss2022conditional,zhu2022generative}. Quantum Boltzmann Machines (QBMs)~\cite{amin2018quantum} represent Gibbs-type quantum states through parameterized Hamiltonians. The original Quantum Variational Autoencoder (QVAE)~\cite{khoshaman2018quantum} uses a QBM as its latent generative process, while later QVAE variants use different latent-state and encoder--decoder constructions~\cite{gao2020high,li2022scalable,wang2025quantum}.

\subsection{Motivations}

	While these models have achieved notable progress in their respective applications, their mechanisms for mixed-state generation remain model-specific. Under the minimax setting analyzed in~\cite{braccia2021enhance}, the QGAN objective is biased toward pure-state optima, making the training objective poorly aligned with convergence to a desired mixed-state target. In its original form, QCBM only generates classical bit strings sampled from a measurement distribution~\cite{benedetti2019generative,liu2018differentiable}. QBM represents Gibbs states of a parameterized Hamiltonian, which confines the generated states to a Hamiltonian-defined family and limits its generality~\cite{amin2018quantum,zoufal2021variational}. The original QVAE relies on a QBM as its latent generative process, while its architectural variants still produce the target state in one generation stage~\cite{khoshaman2018quantum,wang2025quantum}. More broadly, these models generate the target without a prescribed sequence of intermediate states, so they must bridge the entropy gap from initialization to the target in one step, which makes training and generation difficult. These limitations call for a new quantum generative paradigm for mixed states, one that supports stable supervised training, multistep generation, and broad applicability.

	Diffusion-style generation offers such a paradigm. Instead of learning a single generation map, it builds a forward trajectory that progressively corrupts the target with noise and trains reverse steps to reconstruct it~\cite{ho2020denoising,song2020denoising}. This paradigm assigns each reverse step a fixed supervised target that stabilizes optimization, spreads entropy reduction across timesteps so that each step solves an easier subproblem, and may improve robustness under quantum noise~\cite{parigi2023quantum}. These properties make diffusion-style generation a natural candidate for mixed-state quantum generation.

	However, realizing this idea for mixed quantum states is nontrivial. Since the generated object is a density operator, both the forward process and the learned reverse steps must be physically valid quantum operations. The reverse direction cannot be implemented by an unconstrained denoising map, by directly inverting the forward process, or by applying an ordinary unitary only to the noisy system. Instead, each reverse step must be learned as a quantum channel that reduces the additional mixedness introduced by the forward process while preserving the intrinsic mixedness of the target state. Constructing such a trainable reverse-channel trajectory is the central methodological problem addressed in this work.

\begin{figure}[t]
	\centering
	\includegraphics[width=0.80\linewidth]{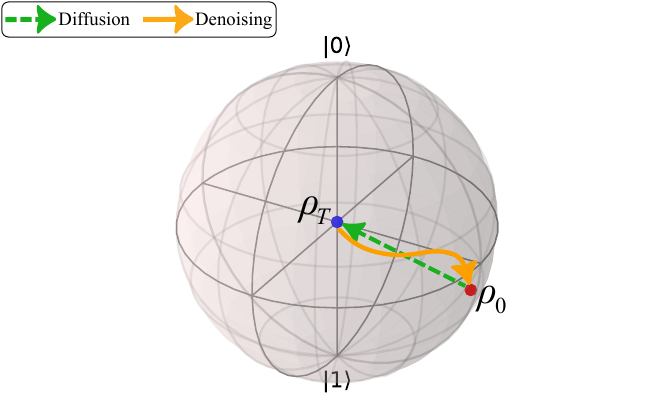}
	\caption{The core idea of the proposed QGDM.}
	\label{fig: main idea}
\end{figure}

\subsection{Contributions}

This paper introduces a quantum generative diffusion model (QGDM) whose forward-reverse structure is grounded in quantum channel theory.
Fig.~\ref{fig: main idea} shows its core idea in a Bloch sphere: any target quantum
state ($\rho_0$) can be transformed into a completely mixed state ($\rho_T$)
through a forward (diffusion) process; subsequently, a trainable backward
(denoising) process can be used to recover the former from the latter. Our
objectives are to a) define a timestep-dependent forward process, b) design a physically valid, trainable backward quantum process, c) theoretically characterize this design, and d) benchmark our model against other models in
pure-, mixed-, and structured-state generation tasks. We aim to make the
following novel and significant contributions to the field of quantum generative
learning:

\begin{enumerate}
	\item We propose a quantum generative diffusion model called QGDM.
	Any target quantum state can be diffused into a completely mixed state by using
	a forward process. Then, using a trainable backward process, the former can be
	recovered from the latter. The forward process employs a depolarizing channel\cite{nielsen2010quantum,wilde2013quantum}.
	To implement a physically valid backward process, we realize it as a variational CPTP map via a Stinespring-type dilation with a trainable joint unitary and a partial trace over an auxiliary register\cite{wilde2013quantum,watrous2018theory}. The resulting channel is non-unitary and supports full-rank mixed-state outputs. Additionally,
	to reduce the number of parameters, we use a parameter-sharing strategy and
	include temporal information as an input, allowing the backward processes at
	all timesteps to share the same parameters.

	\item We introduce a Resource-efficient version of QGDM (RQGDM) to minimize the
	number of auxiliary qubits required in the backward process. Its core idea is to
	utilize a parameterized quantum circuit to extract a low-dimensional
	representation of quantum data from its original high-dimensional Hilbert space.
	This compression is crucial for the backward process since the low-dimensional
	representation uses much fewer qubits than the input state. This
		compression is most useful for low-rank mixed-state generation.
	
	\item We provide a theoretical characterization of this design. First, any exact reverse step must be a non-unital channel fed by a non-maximally-mixed ancilla, so the timestep state must supply the purity that denoising consumes. Second, the diffusion depth and the discarded-register width jointly bound the entropy a reverse trajectory can remove. Third, the per-timestep fidelity losses upper-bound the end-to-end generation error. Finally, the disjoint-register layout removes a passive copy shortcut that would otherwise trap training.
	
	\item We conduct numerical simulations on two complementary benchmarks.
		On random pure- and mixed-state targets, which serve as structure-free
		expressivity tests, QGDM and RQGDM achieve 53.02\% higher fidelity than
		QGAN-based models in mixed-state generation. On structured TFIM Gibbs states
		under gate-level depolarizing noise, QGDM is competitive with the
		Hamiltonian-specialized variational Gibbs state preparation (VGSP) method in the noiseless case and attains the highest mean
		fidelity among the tested methods under gate noise.
\end{enumerate}

We set up open-source codes to facilitate future related
research at https://github.com/ChuangtaoChen/QGDM.

The remainder of this article is structured as follows. Section \ref{section: related works}
briefs the existing literature and the foundational work relevant to quantum generative models.
Section \ref{section: QGDM} presents an in-depth exposition of QGDM.
Section \ref{section: RQGDM} introduces RQGDM.
Section \ref{section: design about f} analyzes the rationale behind the design of the
backward process and presents simulation results.
Section \ref{section: simulations} gives the numerical simulation setup and comparative results.
Finally, Section \ref{section: conclusion} concludes this article.

\section{Preliminary and Related Work}
\label{section: related works}

\subsection{Preliminary}
We first introduce the necessary quantum information background. For more details, please refer to Ref.~\cite{nielsen2010quantum}. The notations are summarized in Table\,\ref{tab:Notations}.

\begin{table}[t]
	\centering
	\caption{Notations.}
	\begin{tabular}{|l|l|}
		\hline
		Notation & Description \\
		\hline
		$\rho_0$    &  Target Quantum State \\\hline
		$\rho_t$    &  System's Density Matrix at Timestep $t$ \\\hline
		$\tau_t$    &  Timestep Embedding Quantum State at Timestep $t$\\\hline
		$\alpha_t$  &  Noise Parameter Dependent at Timestep $t$\\\hline
		$\bar{\alpha}_t$    &  Accumulated Noise Parameter $\bar{\alpha}_t=\prod_i^t{\alpha_i}$ \\\hline
		$\mathcal{E}(\cdot, \cdot)$  &  Forward Process Function \\\hline
		$f_{\boldsymbol{\Theta}}(\cdot, \cdot)$  &  Backward Process Function Parameterized by $\boldsymbol{\Theta}$ \\\hline
		$\mathcal{T}(\boldsymbol{\omega}, \cdot)$    &  Timestep Embedding Circuit Parameterized by $\boldsymbol{\omega}$ \\\hline
		$\mathcal{U}$ & Uniform Distribution\\\hline
		$U(\boldsymbol{\theta})$    &  Denoising Circuit Parameterized by $\boldsymbol{\theta}$ \\\hline
		$\text{tr}(\cdot)$    &  Trace Operation \\\hline
		$\text{tr}_{\textit{B}}(\cdot)$    &  Partial Trace Operation over Subsystem \textit{B}\\\hline
		$N$    &  Number of Qubits in the Target State \\\hline
		$N_{\tau}$  &  Number of Qubits in the Timestep Embedding Circuit \\\hline
		$d$     &  Dimension of Hilbert Space \\\hline
		$\lambda$     &  Hyperparameter in Loss Function \\\hline
		$A^{\dagger}$     &  Hermitian Conjugate of the Operator $A$ \\\hline
		$\mathcal{L}_{t}$     &  Loss Function at Timestep $t$ \\\hline
		
	\end{tabular}
	\label{tab:Notations}
\end{table}

\subsubsection{Pure Quantum State}
In quantum computing, a state of an $N$-qubit can be represented by a vector in a complex Hilbert space $\mathbb{C}^{d}$ with unit length, denoted as $|\psi\rangle \in \mathbb{C}^{d}$, where the \textit{ket} notation $|\rangle$ denotes a column vector and $d=2^N$ is the dimension of Hilbert space. The \textit{bra} notation $\langle\psi| = |\psi\rangle^{\dagger}$ represents a row vector, with $\dagger$ denoting the conjugate transpose. The evolution of a pure state can be represented by: $|\psi'\rangle = U|\psi\rangle$, where $U$ is a unitary operator satisfying $UU^{\dagger} = \mathbb{I}$, and $\mathbb{I}$ is the identity matrix.

\subsubsection{Mixed Quantum State} A mixed state of a quantum system is a probabilistic mixture of pure states. It can be represented by a density matrix (or density operator): $\rho=\sum_{i}p_i|\psi_i\rangle\langle\psi_i|$, where the ensemble of pure states $\{p_i, |\psi_i\rangle\}$ indicates that the system is in state $|\psi_i\rangle$ with a probability of $p_i$. This matrix is particularly useful in contexts where the exact state of a system is not known or when considering the subsystems of entangled pairs. The evolution of a mixed state is described by $\rho'=U\rho U^{\dagger}$. More general mixed-state dynamics, including open-system and noisy evolution, are described by completely positive and trace-preserving (CPTP) maps.

\subsubsection{Tensor Product} 
The Hilbert space of a combined system, composed of two (or more) different quantum systems, is the tensor product of the Hilbert spaces of the subsystems. Specifically, if the states of the subsystems are $|\psi_A\rangle$ and $|\psi_B\rangle$ for pure states, or $\rho_A$ and $\rho_B$ for mixed states, the state of the composite system can be represented as $|\psi_A\rangle \otimes |\psi_B\rangle$ or $\rho_A\otimes\rho_B$, respectively. The tensor product operation allows us to describe the complete state of the composite system through the individual states of its components. 

\subsubsection{Partial Trace} Suppose that we have two quantum systems, \textit{A} and \textit{B}. Their composite system state is represented by the density matrix $\rho_{AB}$. The partial trace operation yields a reduced density matrix for subsystem \textit{A}, calculated by $\rho_A = \text{tr}_B(\rho_{AB})$. Its definition is $\text{tr}_B\left(|a_1\rangle\langle a_2| \otimes |b_1\rangle\langle b_2|\right) = |a_1\rangle\langle a_2|\text{tr}\left(|b_1\rangle\langle b_2|\right)$, where $|a_1\rangle$ and $|a_2\rangle$ are any two vectors in \textit{A}, while $|b_1\rangle$ and $|b_2\rangle$ are any two vectors in \textit{B}.

\subsection{Quantum Generative Models}

Quantum generative models, a pivotal sector in quantum machine learning, encompass various models including QCBM \cite{benedetti2019generative}, QBM \cite{amin2018quantum}, QVAE \cite{khoshaman2018quantum}, and QGAN \cite{lloyd2018quantum, dallaire2018quantum}.

QCBM uses quantum circuits to generate a classical bit string from specific probability distributions \cite{benedetti2019generative}. This model has substantial advanced \cite{liu2018differentiable, du2020expressive, zhu2019training, coyle2020born}, finding its applications in various fields, $e.g.$, Monte Carlo simulations\cite{kiss2022conditional}, financial data generation\cite{coyle2021quantum}, and joint distribution learning\cite{coyle2021quantum, zhu2022generative}.
QBM functioning as a quantum counterpart of the classical Boltzmann Machine \cite{fahlman1983massively,ackley1985learning},
uses a quantum device to prepare Boltzmann distributions, by replacing the units in the Boltzmann Machine with qubits and substituting the energy function with a parameterized Hamiltonian. 
QBM was first demonstrated in \cite{amin2018quantum}, has its enhancements \cite{kieferova2017tomography, zoufal2021variational}.
Khoshaman et al.~\cite{khoshaman2018quantum} introduce the original QVAE, in which the latent generative process is implemented by a QBM, with later developments including annealer-based and gate-based QVAE variants~ \cite{gao2020high,li2022scalable}.
In the realm of QGANs, Lloyd et al. \cite{lloyd2018quantum} develope a comprehensive taxonomy, classifying them by the quantum or classical nature of their components. 
This was empirically supported by Dallaire-Demers and Killoran \cite{dallaire2018quantum}, who demonstrated the efficacy of QGAN in generating the desired quantum data. Since then, the field has continuously explored applications in quantum state generation \cite{benedetti2019adversarial,braccia2021enhance,niu2022entangling}, image generation \cite{huang2021experimental,silver2023mosaiq,tsang2023hybrid}, and discrete data generation\cite{situ2020quantum,zeng2019learning,chaudhary2023towards}. For an in-depth review of these quantum generative models, please refer to \cite{tian2023recent}.

During our research, several studies \cite{parigi2023quantum,zhang2024generative,cacioppo2023quantum} have proposed quantum analogs of classical diffusion models, but their tasks and physical mechanisms differ from the present mixed-state channel-generation setting. Parigi et al.~\cite{parigi2023quantum} discuss strategies and tools for developing quantum diffusion models, but do not provide a concrete mixed-state generation benchmark. Zhang et al.~\cite{zhang2024generative} propose QuDDPM for generating individual pure states from a target distribution, whereas QGDM targets density-operator generation and uses a shared-parameter reverse channel across timesteps. Cacioppo et al.~\cite{cacioppo2023quantum} use a classical approach for their diffusion process, while the denoising process is performed via a trainable quantum circuit for the denoising task. However, our method is fully quantum-mechanical, encompassing both diffusion and denoising processes.

\section{Quantum Generative Diffusion Model (QGDM)}
\label{section: QGDM}

\subsection{General Framework}
\begin{figure*}[t]
	\centering
	\includegraphics[width=0.80\textwidth]{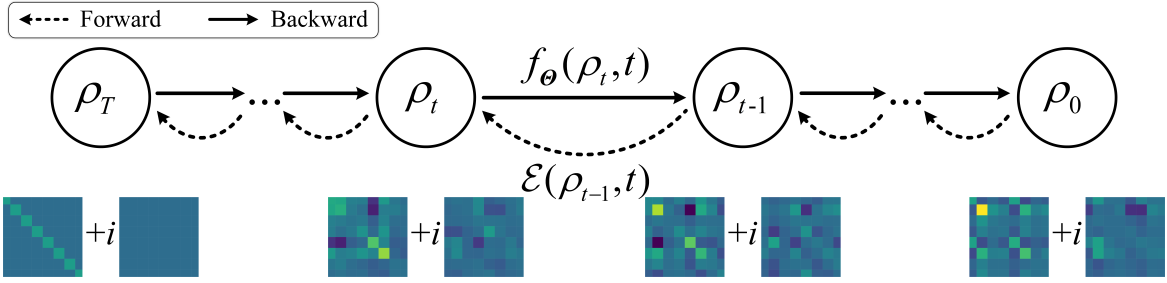}
	\caption{The framework of the proposed Quantum Generative Diffusion Model (QGDM).}
	\label{fig:general_framework}
\end{figure*}

Fig.\,\ref{fig:general_framework} shows a framework of QGDM.
It includes a forward process (or diffusion one) $\mathcal{E}(\rho_{t-1},t)$ and a backward process (or denoising one) $f_{\boldsymbol{\Theta}}(\rho_t,t)$. $\mathcal{E}(\rho_{t-1},t)$ takes an $N$-qubit quantum state $\rho_{t-1}$ and a timestep scalar $t$ as input, producing a quantum state $\rho_{t}$ with added noise. 
Similar to classical diffusion models, in QGDM, a ``timestep" refers to a discrete step in the model's processes where noise is sequentially added to (forward process) or subtracted from a quantum state (backward process).
In this paper, $\mathcal{E}(\rho_{t-1},t)$ is modeled as a depolarizing channel. When $t=0$, $\rho_0$ represents the target quantum state. In contrast, at $t=T$, $\rho_T$ is a completely mixed state. The denoising process employs a model $f_{\boldsymbol{\Theta}}(\rho_{t},t)$ with trainable parameters $\boldsymbol{\Theta}$, which denoises the noisy input quantum state $\rho_{t}$, producing a cleaner quantum state $\rho_{t-1}$. 
After applying the learned map $f_{\boldsymbol{\Theta}}(\cdot,\cdot)$ for $T$ steps from $\rho_T$, the model generates an estimate of the target state $\rho_0$.

\textbf{Remark 1:}
Let us discuss the difference between the notations used in this paper and those used in classical diffusion model papers \cite{ho2020denoising,dhariwal2021diffusion,luo2022understanding}. In quantum information, the quantum operation formalism $\rho' = \mathcal{E}(\rho)$ is a powerful tool for describing the evolution of quantum states \cite{nielsen2010quantum}, where $\mathcal{E}$ represents a quantum operation, and $\rho$ and $\rho'$ are the initial and final states, respectively. This formalism is especially useful for describing open quantum systems or the stochastic evolution of quantum states. In this paper, we utilize the quantum operation formalism, denoting forward and backward processes as $\mathcal{E}(\rho_{t-1}, t)$ and $f_{\boldsymbol{\Theta}}(\rho_t, t)$, respectively. These notations do not explicitly include the output term.
In contrast, in classical diffusion models \cite{sohl2015deep,ho2020denoising}, forward and reverse distributions are defined as $q(x_t|x_{t-1})$ and $p_{\boldsymbol{\theta}}(x_{t-1}|x_t)$. These notations explicitly include the output term ($i.e.$, $x_t$ in $q(x_t|x_{t-1})$ and $x_{t-1}$ in $p_{\boldsymbol{\theta}}(x_{t-1}|x_t)$). The difference between the notations in our paper and those in classical diffusion model papers can be summarized as follows.
In the classical case, the evolution process of a sample ($x_{t-1}$ or $x_t$) is probabilistic. Therefore, the forward or backward process uses conditional probability notation. In the quantum case, the system's state is described by a density matrix. The evolution of the density matrix is deterministic, but the density matrix itself represents a probabilistic quantum system. Thus, we use the quantum operation formalism to describe the forward and backward processes.
This distinction between quantum and classical systems aids readers who may not be familiar with quantum computation in understanding our notations.

\subsection{Forward Process}

In the natural world, diffusion is governed by non-equilibrium thermodynamics, driving systems toward equilibrium \cite{de2013non}. As a system undergoes evolution, its entropy increases until it reaches the maximum at equilibrium, indicating that there is no further evolution.
The classical diffusion model draws inspiration from this diffusion mechanism \cite{sohl2015deep}.
In quantum information theory, the depolarizing channel is considered a ``worst-case scenario'' \cite{wilde2013quantum}. Under its influence, a quantum state gradually loses its original information, eventually transforming into a completely mixed state. This state is characterized by the maximum entropy, representing the pinnacle of randomness in quantum systems. At this point, our uncertainty about the system has also reached its maximum.

An $N$-qubit completely mixed state is mathematically represented as $\mathbb{I}/{2^N}= \sum\limits_{i} {\frac{1}{{{2^N}}}} \left| i \right\rangle \left\langle i \right|$. This denotes that each computational basis state $\left| i \right\rangle $ occurs with an equal probability of $\frac{1}{2^N}$.
This uniform probability distribution across all computational bases shows the state to be maximally random, similar to the uniform distribution in classical data.
At timestep $t$, the diffusion process of QGDM adds white noise to an $N$-qubit quantum state $\rho_{t-1}$:
\begin{equation}
	\begin{aligned}
		\label{equ:diffusion processing}  
		\rho_{t}&=\mathcal{E}\left(\rho_{t-1},t\right)\\
		&=(1-\alpha_t) \mathbb{I}/d+\alpha_t \rho_{t-1},
	\end{aligned}
\end{equation}
where $d=2^N$ denotes the dimension of Hilbert space and $\alpha_t\in [0, 1]$ is a timestep-dependent scalar. 
The completely mixed state $\mathbb{I}/d$ is the maximally mixed, maximum-entropy state. In this generative setting, it plays the role of white noise, namely a structureless reference state toward which the forward process drives the target. Eq.~(1) can be interpreted as a depolarizing step that keeps the input state $\rho_{t-1}$ with weight $\alpha_t$ and replaces it by the white-noise state $\mathbb{I}/d$ with weight $1-\alpha_t$. Equivalently, each one-step forward state is a generalized Werner-type mixture of $\rho_{t-1}$ and $\mathbb{I}/d$. By Theorem~1, the state at any timestep can also be written as a generalized Werner-type mixture of the target state $\rho_0$ and white noise. As $t$ increases, the weight on $\mathbb{I}/d$ grows, and the original quantum information is progressively diluted.
This terminology follows the generalized Werner-state construction~\cite{werner1989quantum}. The progressive dilution indicates an irreversible loss of information in the quantum system; consequently, the depolarizing channel is non-unitary. By setting an appropriate sequence of $\{\alpha_{t}\}$, a sequence of depolarizing channels acting on the quantum state $\rho_0$ can eventually yield the output $\rho_T\approx \mathbb{I}/d$.

\subsubsection{Efficient Quantum State Transition to Arbitrary Timestep}
Eq.\,(\ref{equ:diffusion processing}) mathematically describes a single-step diffusion process. It is inefficient to repeatedly apply $\mathcal{E}$ to obtain a specific timestep's quantum state $\rho_t$.
To enhance the transition efficiency from the initial state $\rho_0$ to a specific state $\rho_t$, we introduce Theorem 1.

\textbf{Theorem 1:}
\textit{For any $t\in\{1,\dots,T\}$, the relationship between $\rho_0$ and $\rho_t$ can be described by:
\begin{equation}
	\label{eq: one step diffusion}
	\begin{aligned}
		\rho_t=\left(1-\bar{\alpha}_{t}\right) \mathbb{I}/d+\bar{\alpha}_{t} \rho_0,
	\end{aligned}
\end{equation}
where $\bar{\alpha}_{t}=\prod\limits_{i = 1}^t \alpha_i$. Thus, $\bar{\alpha}_{t} \le \bar{\alpha}_{t-1}$ since $\alpha_i\in [0,1],\forall i$.}

\smallskip
\noindent\textit{Proof.}\; A derivation is provided in the Supplementary Material.\hfill$\square$

Eq.\,(\ref{eq: one step diffusion}) allows $\rho_t$ to be obtained directly from $\rho_0$ at any chosen timestep.

\subsubsection{Noise Schedule}
The choice of ${\alpha_t}$ must ensure that $\bar{\alpha}_T \to 0$ and ${\alpha _t} \in \left[ {0,1} \right]$ for all $t$. We use the cosine noise schedule from the classical diffusion model \cite{nichol2021improved} to select $\alpha_t$:

\begin{equation}
	\label{eq:beta generation}
	\alpha_t=\frac{\bar{\alpha}_t}{\bar{\alpha}_{t-1}},
\end{equation}
where
\begin{equation}
	\bar{\alpha}_t=\frac{g(t)}{g(0)}, \quad g(t)=\cos \left(\frac{t / T+s}{1+s} \cdot \frac{\pi}{2}\right)^2,
\end{equation}
with $s$ being a small offset hyperparameter. 
We note that there are other noise schedules, such as the linear noise schedule \cite{ho2020denoising} and learned one \cite{kingma2021variational}. Exploring the impacts of other noise schedules on QGDM's performance, or designing specific noise schedules for QGDM, is an interesting research direction, which we leave for future work.

\begin{figure}[t]
	\centering
	\includegraphics[width=0.75\linewidth]{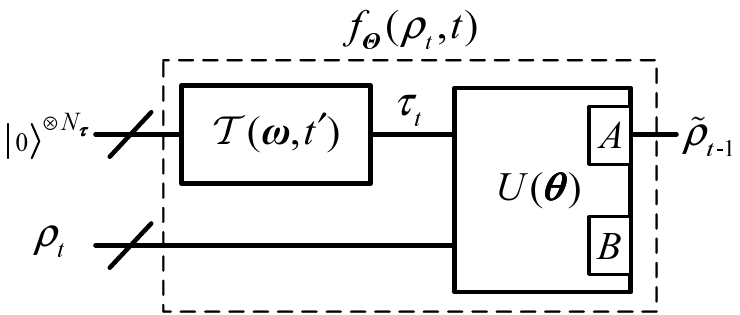}
	\caption{The denoising process framework, $f_{\boldsymbol{\Theta}}(\rho_t,t)$, is composed of a timestep embedding circuit $\mathcal{T}(\boldsymbol{\omega},t')$ and a denoising circuit $U(\boldsymbol{\theta})$, where $\boldsymbol{\Theta}=\{\boldsymbol{\omega},\boldsymbol{\theta}\}$.}
	\label{fig:original}
\end{figure}

\subsection{Backward Process}
\label{subsection: method original}

The objective of the backward process in QGDM is to restore $\rho_0$ from $\rho_T$. Before presenting the architecture, we first address two key considerations.

Firstly, because the depolarizing forward step has no exact CPTP inverse, the backward process cannot be realized as an ordinary unitary acting only on the noisy system; it must be modeled as a trainable non-unitary quantum channel.
Secondly, modeling $T$ distinct denoising steps separately would require a large number of trainable parameters. To keep the model compact, we let all denoising steps share the same parameters by conditioning on the timestep $t$.

To endow the denoising process with the above characteristics, we realize non-unitary evolution by tracing out part of an auxiliary subsystem, while embedding temporal information via a parameterized circuit. Our design for the denoising process $f_{\boldsymbol{\Theta}}(\rho_t, t)$ is illustrated in Fig.\,\ref{fig:original}. It consists of two modules: a timestep embedding circuit $\mathcal{T}(\boldsymbol{\omega}, t')$ (with $t' = t\pi/T$) acting on an auxiliary register of $N_{\tau}$ qubits, and a denoising circuit $U(\boldsymbol{\theta})$ acting on the composite system. The trainable parameters are $\boldsymbol{\Theta} = \{\boldsymbol{\omega}, \boldsymbol{\theta}\}$.

Specifically, the timestep embedding circuit produces a state $\tau_t$ that carries temporal information. The composite system $\tau_t \otimes \rho_t$ is then processed by the denoising circuit $U(\boldsymbol{\theta})$. Afterward, the qubits are regrouped into register $A$ ($N$ qubits) and register $B$ ($N_{\tau}$ qubits). We trace out register $B$ to obtain the predicted denoised state $\tilde{\rho}_{t-1}$.


\subsubsection{Timestep Embedding Circuit}
A critical consideration when designing the denoising process is how to encode the temporal information into a quantum state. This is a classical-to-quantum encoding challenge, where such methods as qubit encoding \cite{stoudenmire2016supervised,grant2018hierarchical} or amplitude encoding \cite{rebentrost2014quantum} could be used to embed $t$ into a quantum system. However, the qubit encoding method only maps the timestep sequence uniformly onto a ring on the Bloch sphere, which lacks expressiveness and flexibility. The amplitude encoding method requires a number of gates that is exponential in the number of qubits \cite{larose2020robust}. To enhance the expressiveness and flexibility of timestep embedding while maintaining an appropriate gate complexity, this work utilizes the quantum embedding method \cite{lloyd2020quantum} to obtain a timestep embedding state $\tau_t$ with $N_{\tau}$ qubits, $i.e.$,

\begin{equation}
	\label{eq:time embedding state}
	\tau_t = \mathcal{T}(\boldsymbol{\omega}, t')\left(\left| {\rm{0}} \right\rangle {\left\langle {\rm{0}} \right|}\right)^{ \otimes N_{\tau}}\mathcal{T}^{\dagger}(\boldsymbol{\omega}, t'),
\end{equation}
where scalar $t'=t\pi/T$ is the mapping of timestep $t$ into the range [0, $\pi$], and $\boldsymbol{\omega}$ denotes the trainable parameters in the timestep embedding circuit. 

\begin{figure}[h]
	\centering
	\includegraphics[width=0.8\linewidth]{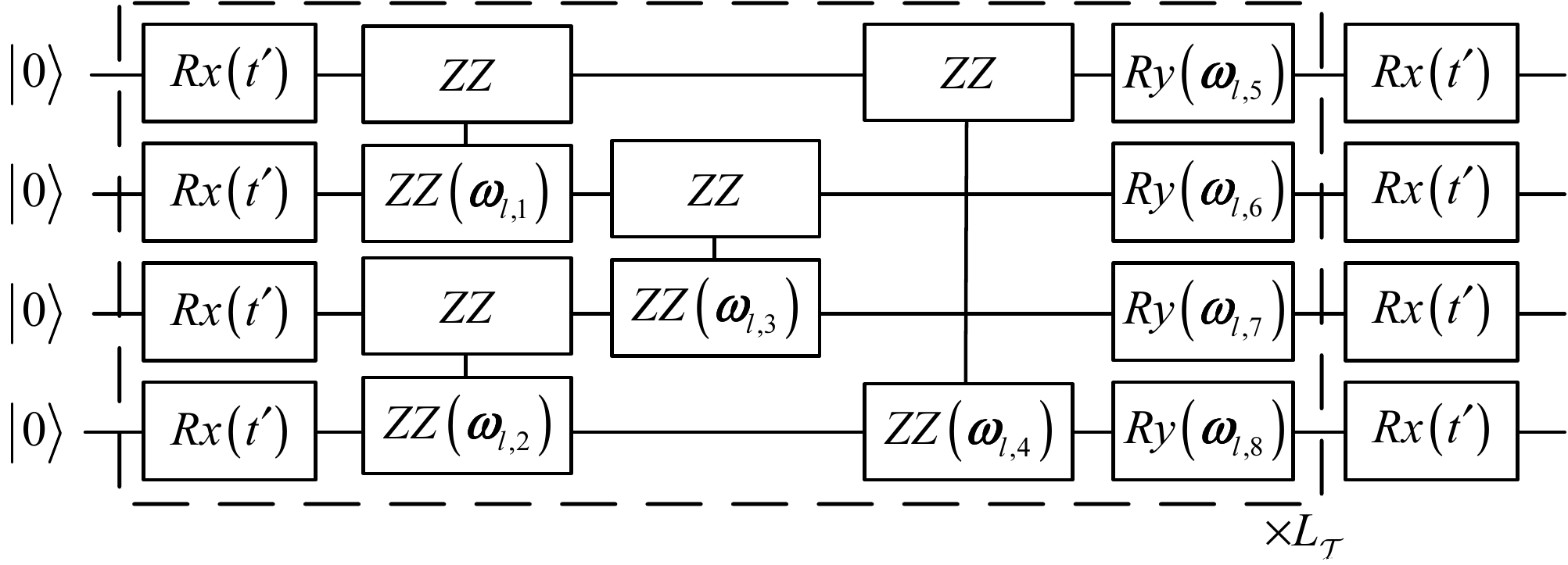}
	\caption{The conventional ansatz \cite{lloyd2020quantum} is used for our timestep embedding with trainable parameter $\boldsymbol{\omega}$, and given parameter $t' = t\pi/T$. The circuit in the dashed lines is repeated for $L_{\mathcal{T}}$ times. Where $l\in\{1,\dots,L_{\mathcal{T}}\}$.}
	\label{fig:time embedding circuit}
\end{figure}

The architecture of a timestep embedding circuit with $N_{\tau}=4$ qubits is shown in Fig.\,\ref{fig:time embedding circuit}. The circuit includes single-qubit gates $Rx$ and $Ry$, and the two-qubit gate $ZZ(\phi)=\exp{(-i\phi (Z\otimes Z) /2)}$.

\subsubsection{Denoising Circuit}

The denoising circuit $U(\boldsymbol{\theta})$ acts on the composite system $\tau_t\otimes \rho_t$. Subsequently, all qubits are regrouped into subsystems \textit{A} and \textit{B}, where they possess $N$ and $N_{\tau}$ qubits, respectively. By tracing out B, we obtain the predicted output:

\begin{equation}
	\label{eq:denoising process}
	\begin{aligned}
		{{{\tilde \rho }_{{\rm{t}} - 1}}}
		&=f_{\boldsymbol{\Theta}}(\rho_t,t)\\
		&=\text{tr}_\text{\textit{B}}\left( U(\boldsymbol{\theta})\left(\tau_t \otimes \rho_t\right)U^{\dagger}(\boldsymbol{\theta})\right).
	\end{aligned}
\end{equation}


Using subsystem \textit{B} to output ${\tilde \rho }_{t-1}$ seems intuitive. However, such a design would result in model failure. This failure is caused by the slight difference between $\rho_t$ and ground truth $\rho_{t-1}$, which poses a training challenge. Rather than producing the intended output, the optimizer often merely copies its input. The next Section provides the detailed examination of this issue, including numerical simulations to demonstrate the concept.

\begin{figure}[h]
	\centering
	\includegraphics[width=1.0\linewidth]{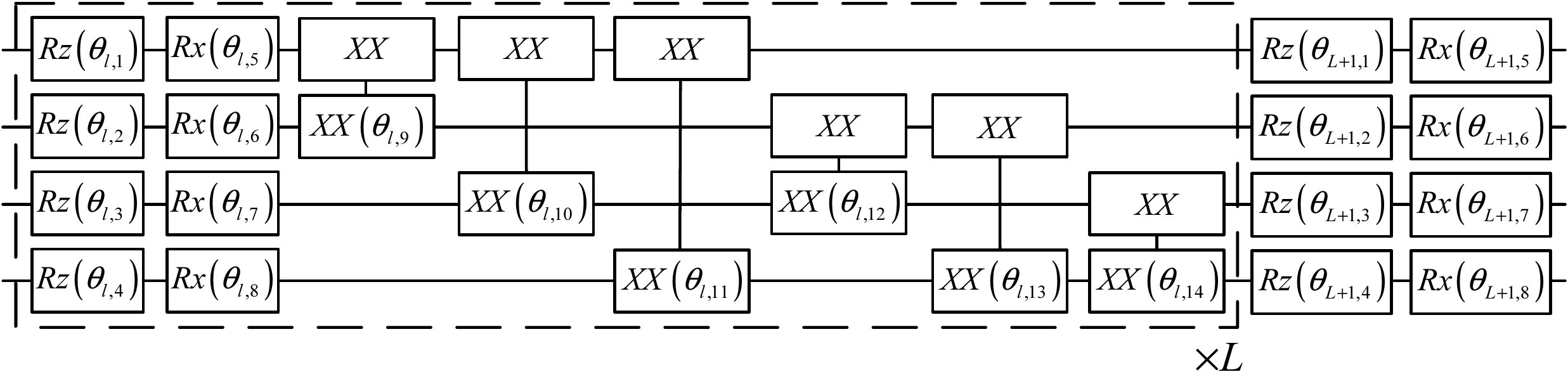}
	\caption{The architecture of a 4-qubit denoising circuit $U(\boldsymbol{\theta})$, where $\boldsymbol{\theta}$ denotes trainable parameters. To enhance its learning capability, the dashed-boxed part is repeated $L$ times. $l\in \{1,\dots,L\}$.}
	\label{fig:PQC}
\end{figure}

The architecture of a 4-qubit denoising circuit $U(\boldsymbol{\theta})$ is shown in Fig.\,\ref{fig:PQC}. The circuit includes single-qubit gates $Rz$ and $Rx$, and the two-qubit gate $XX(\phi)=\exp{(-i\phi (X\otimes X) /2)}$. 


\subsection{Training and Generation}
The training objective of QGDM is to maximize the quantum fidelity between the predicted denoised state $\tilde\rho_{t-1}$ and the ground truth $\rho_{t-1}$. The loss function is defined as follows:
\begin{equation}
	\label{eq: loss function1}
	\min _{\boldsymbol{\Theta}} \mathcal{L}=\mathcal{L}_0+\lambda \mathbb{E}_{t \in \mathcal{U}(2,\dots,T)}\left[\mathcal{L}_{t-1}\right],
\end{equation}
where $\lambda$ is a hyperparameter used to balance the losses $\mathcal{L}_0$ and $ \mathbb{E}_{t \in \mathcal{U}(2,\dots,T)}\left[\mathcal{L}_{t-1}\right]$, and we find that setting a small $\lambda$ improves the generative effect of QGDM. $\mathcal{U}(2,\dots,T)$ represents a uniform distribution that gives values from 2 to $T$. We have:
\begin{equation}
	\label{eq: loss function2}
	{\mathcal{L}_{t-1}} = 1 - F\left(\rho _{t-1},{\tilde \rho }_{t-1}\right),
\end{equation}
where $F(\cdot, \cdot)$ is the quantum fidelity function \cite{nielsen2010quantum,wilde2013quantum}, used to measure the distance between two quantum states $\rho$ and $\sigma$:
\begin{equation}
	\label{eq: loss function3}
	F(\rho, \sigma) = \left(\operatorname{tr} \sqrt{\sqrt{\rho} \sigma \sqrt{\rho}}\right)^2.
\end{equation}
This objective is directly related to the Bures distance $D_{\mathrm{B}}(\rho,\sigma)=\sqrt{2-2\sqrt{F(\rho,\sigma)}}$~\cite{hubner1992explicit}, since it is a monotone function of the fidelity. Thus, maximizing fidelity is equivalent to minimizing the Bures distance for the denoising target.

QGDM training is summarized in Algorithm~\ref{alg:training}. 
In each iteration, we compute $\mathcal{L}_0$ and the expectation $\mathbb{E}_{t \in \mathcal{U}(2,\dots,T)}[\mathcal{L}_{t-1}]$ to update the trainable parameters until the training loss converges or the maximum number of iterations is reached. 
The expected value is estimated by using the Monte Carlo approach.

Once the parameters are trained, we can iteratively obtain the target state $\rho_0$ from the completely mixed state $\rho_T$ by using $\boldsymbol{\omega}^{}$ and $\boldsymbol{\theta}^{}$. $\rho_0$ is then generated via Algorithm~\ref{alg:generation}.
The algorithm starts with a completely mixed state and uses the trained denoising process to generate the target state via $T$ steps. Therefore, its complexity is $\mathcal{O}(T)$.
It is important to distinguish the supervised training trajectory from the generation trajectory. During training, the forward process constructs $\rho_0,\rho_1,\ldots,\rho_T$, and the denoiser at timestep $t$ is trained with input $\rho_t$ and target $\rho_{t-1}$. In particular, the final reverse step is supervised by the given target state $\rho_0$, which may itself be mixed. During generation, the learned reverse channels are applied for the fixed sequence $t=T,T-1,\ldots,1$ starting from $\tilde{\rho}_T=\mathbb{I}/d$, and the process stops at $\tilde{\rho}_0$ by construction. Thus, QGDM does not use an adaptive purification criterion and does not continue denoising a mixed target toward an underlying pure component.

\begin{algorithm}[h]
	\caption{QGDM training.}  
	\label{alg:training}
	\begin{algorithmic}[1]   
		\STATE \textbf{Input:} Noise schedule parameters $\{\bar{\alpha}_t\}_{t=1}^{T}$, batch size $\text{BS}$, weight parameter $\lambda$, number of qubits of the target state $N$, number of qubits of the timestep embedding state $N_{\tau}$, total timestep of the diffusion process $T$, maximum iteration number,
		target quantum state $\rho_0$, learning rate $\eta$, completely mixed state $\mathbb{I}/d$.
		\STATE $\boldsymbol{\theta} \sim  \mathcal{U}(0, \pi)$, $\boldsymbol{\omega} \sim \mathcal{U}(0, \pi)$
		\REPEAT
		\STATE $\rho_{1}\leftarrow\left(1-\bar{\alpha}_{\mathrm{1}}\right) \mathbb{I}/d+\bar{\alpha}_{\mathrm{1}} \rho_0$
		\STATE $\tau_1 \leftarrow \mathcal{T}(\boldsymbol{\omega}, \pi/T)\left(\left| {\rm{0}} \right\rangle {\left\langle {\rm{0}} \right|}\right)^{ \otimes N_{\tau}}\mathcal{T}^{\dagger}(\boldsymbol{\omega}, \pi/T)$
		\STATE $\tilde \rho_{0} \leftarrow \text{tr}_\text{\textit{B}}\left( U(\boldsymbol{\theta})\left(\tau_1 \otimes \rho_1\right)U^{\dagger}(\boldsymbol{\theta})\right)$
		\STATE $\mathcal{L}_0\leftarrow 1-F(\rho_0, \tilde \rho_0 )$
		\STATE $\mathcal{L} \leftarrow 0$
		\FOR{$i$ from 1 to $\text{BS}$}
		\STATE  Sample a unique timestep $t$ from the uniform distribution $\mathcal{U}(2,\dots,T)$
		\STATE $\rho_{t}\leftarrow\left(1-\bar{\alpha}_{\mathrm{t}}\right) \mathbb{I}/d+\bar{\alpha}_{\mathrm{t}} \rho_0$
		\STATE $\tau_t \leftarrow \mathcal{T}(\boldsymbol{\omega}, t\pi/T)\left(\left| {\rm{0}} \right\rangle {\left\langle {\rm{0}} \right|}\right)^{ \otimes N_{\tau}}\mathcal{T}^{\dagger}(\boldsymbol{\omega}, t\pi/T)$
		\STATE $\tilde \rho_{t-1} \leftarrow \text{tr}_\text{\textit{B}}\left( U(\boldsymbol{\theta})\left(\tau_t \otimes \rho_t\right)U^{\dagger}(\boldsymbol{\theta})\right)$
		\STATE $\rho_{t-1}\leftarrow\left(1-\bar{\alpha}_{\mathrm{t-1}}\right) \mathbb{I}/d+\bar{\alpha}_{\mathrm{t-1}} \rho_0$
		\STATE $\mathcal{L} \leftarrow \mathcal{L} + 1-F\left(\rho_{t-1}, \tilde \rho_{t-1}\right)$
		\ENDFOR
		\STATE $\mathcal{L} \leftarrow \mathcal{L}_0 + \lambda\frac{1}{\text{BS}}\mathcal{L}$
		\STATE $\boldsymbol{\omega}\leftarrow \boldsymbol{\omega} - \eta \nabla_{\boldsymbol{\omega}}\mathcal{L};\boldsymbol{\theta} \leftarrow \boldsymbol{\theta} - \eta \nabla_{\boldsymbol{\theta}}\mathcal{L}$
		\UNTIL{$\mathcal{L}$ converges or the number of iterations reaches the maximum}
		\STATE \textbf{Output:} Optimal parameters $\boldsymbol{\omega}^{*}, \boldsymbol{\theta}^{*}$ 
	\end{algorithmic}  
\end{algorithm}  

\begin{algorithm}[h]
	\caption{QGDM generation.}  
	\label{alg:generation}
	\begin{algorithmic}[1] 
		\STATE \textbf{Input:} Number of qubits of the timestep embedding state $N_{\tau}$, total timestep of the diffusion process $T$, the optimal parameters $\boldsymbol{\omega}^{*}, \boldsymbol{\theta}^{*}$, completely mixed state $\rho_T=\mathbb{I}/d$.
		\STATE Initialize $\tilde \rho$ as completely mixed state: $\tilde \rho \leftarrow \rho_T$  
		\FOR{$t$ from $T$ to 1}
		\STATE $t'\leftarrow t\pi/T$
		\STATE $\tau_t\leftarrow \mathcal{T}(\boldsymbol{\omega}^{*}, t')\left(\left| {\rm{0}} \right\rangle {\left\langle {\rm{0}} \right|}\right)^{ \otimes N_{\tau}}\mathcal{T}^{\dagger}(\boldsymbol{\omega}^{*}, t')$
		\STATE Update the generation state: $\tilde \rho\leftarrow \text{tr}_\text{\textit{B}}\left( U(\boldsymbol{\theta}^{*})\left(\tau_t \otimes \tilde \rho\right)U^{\dagger}(\boldsymbol{\theta}^{*})\right)$ using the previous value of $\tilde \rho$.
		\ENDFOR
		\STATE \textbf{Output:} The final generation state $\tilde \rho_0 = \tilde \rho$
	\end{algorithmic}  
\end{algorithm}

\section{Resource-efficient Quantum Generative Diffusion Model (RQGDM)}
\label{section: RQGDM}

The number of qubits required for the QGDM denoising process is $N_{\tau}+N$. To reduce the number of auxiliary qubits, we propose a new design for the denoising process $f_{\boldsymbol{\Theta}}(\rho_t, t)$. We refer to this prototype as RQGDM.

\begin{figure}[h]
	\centering
	\includegraphics[width=0.8\linewidth]{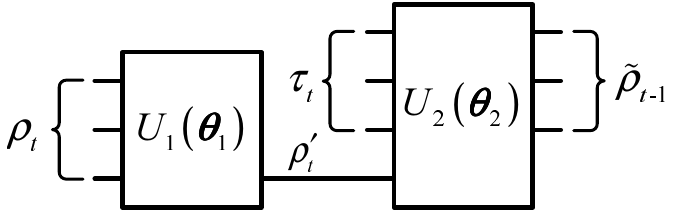}
	\caption{The backward process design in Resource-efficient Quantum Generative Diffusion Models (RQGDM).}
	\label{fig:method like AE}
\end{figure}

Inspired by quantum data compression methods \cite{wan2017quantum,romero2017quantum}, the core idea of RQGDM is to first use a parameterized quantum circuit to learn a low-dimensional representation of quantum data from its original high-dimensional Hilbert space for the denoising process. This low-dimensional representation uses fewer qubits than the input state (e.g., one qubit in our approach), thereby reducing the total number of qubits required for the denoising process. In RQGDM, only $N+1$ qubits are needed to generate an $N$-qubit quantum state, which enhances its efficiency in terms of resource utilization. Fig.\,\ref{fig:method like AE} illustrates the denoising process circuit for a 3-qubit target state using RQGDM. 
The polluted state $\rho_t$ is compressed by a parameterized circuit $U_1(\boldsymbol{\theta}_1)$, with $\boldsymbol{\theta}_1$ as its trainable parameters, condensing its information into the final qubit. This qubit, containing concentrated information, is then combined with $\tau_{t}$ to form a composite system. Another circuit, $U_2(\boldsymbol{\theta}_2)$, processes this expanded state across $N+1$ qubits. After applying $U_2(\boldsymbol{\theta}_2)$, the system's last qubit is discarded, leaving the reduced density matrix of the remaining $N$ qubits as the predicted output state ${\tilde \rho }_{{\rm{t}} - 1}$:
\begin{equation}
	\begin{aligned}
		\tilde{\rho}_{t-1}={\rm{t}}{{\rm{r}}_4}\left[ {U_2\left( {{\boldsymbol{\theta} _2}} \right)\left( {{\tau _t} \otimes \rho_t^{'}} \right){U_2^\dag }\left( {{\boldsymbol{\theta} _2}} \right)} \right],   
	\end{aligned}
\end{equation}
where $\tau_t$ is given by (\ref{eq:time embedding state}) and $\rho_t^{'}$ is calculated by tracing out the first two qubits of the output from the $U_1$ circuit: 
\begin{equation}
	\rho_t^{'}={\rm{t}}{{\rm{r}}_{12}}\left( {U_1\left( {{\boldsymbol{\theta} _1}} \right){\rho _t}{U_1^\dag }\left( {{\boldsymbol{\theta} _1}} \right)} \right).
\end{equation}

Therefore, the trainable parameters for the denoising process $f_{\boldsymbol{\Theta}}(\rho_t,t)$ of RQGDM are $\boldsymbol{\Theta}=\{\boldsymbol{\omega}, \boldsymbol{\theta} _1, \boldsymbol{\theta} _2\}$. 
Both QGDM and RQGDM share the same diffusion process, training, and generation pipeline, but they differ in the denoising process.
This compression is most useful for low-rank mixed-state targets. As the target rank increases, the compressed register must grow accordingly, reducing the auxiliary-qubit savings. In the full-rank case, where the rank equals the Hilbert-space dimension $d=2^N$, the compressed register must span the full target space, so RQGDM reduces to the standard QGDM construction and provides no auxiliary-qubit reduction.

\section{Theoretical Analysis}
\label{section: design about f}
In this section, we provide a detailed analysis of QGDM's denoising process. Throughout, $S(\rho)=-\operatorname{tr}(\rho\ln\rho)$ denotes the von Neumann entropy. 

\subsection{Necessity of a Non-Unital, Purity-Fed Reverse Process}
\label{subsec: necessity}

A simpler reverse process, such as a mixed-unitary channel that uses neither an ancilla nor a partial trace, is insufficient. Since $0<\alpha_t<1$ and $\rho_{t-1}\neq\mathbb{I}/d$, the forward depolarizing step strictly increases the von Neumann entropy by strict concavity~\cite{nielsen2010quantum}. Thus, any exact reverse step must strictly reduce this entropy. The following proposition shows that, even within the ancilla-assisted construction of Eq.~(\ref{eq:denoising process}), the ancilla purity is a necessary resource rather than a free design choice.

\smallskip
\noindent\textbf{Proposition~1} (Entropy budget of the ancilla).
\textit{Let $0<\alpha_t<1$ and $\rho_{t-1}\neq\mathbb{I}/d$. For any denoising process of the form Eq.~(\ref{eq:denoising process}), where $\tau_t$ is an $N_{\tau}$-qubit ancilla, the entropy removed in one backward step satisfies}
\begin{equation}
	\label{eq: prop1 budget}
	S(\rho_t)-S(\rho_{t-1}) \le N_{\tau}\ln 2-S(\tau_t).
\end{equation}
\textit{The forward step makes the left side strictly positive, which forces $S(\tau_t)<N_{\tau}\ln 2$. Thus, the ancilla cannot be maximally mixed. The available entropy-removal budget $N_{\tau}\ln 2-S(\tau_t)$ is maximized when $\tau_t$ is pure ($ S(\tau_t)=0 $).}

\smallskip
\noindent\textit{Proof.}  A detailed proof is provided in the Supplementary Material. It first shows that the forward step strictly increases entropy by using unitality and the strict concavity of the von Neumann entropy. It then applies subadditivity to the joint dilated state to bound the entropy removable in one reverse step, where the discarded register carries at most $N_\tau \ln 2$, yielding $N_\tau \ln 2 - S(\tau_t)$. \hfill$\square$
\medskip

The timestep embedding state must therefore supply purity. In QGDM, the pure timestep state $\tau_t$ plays two roles: it conditions the shared denoiser on the timestep and provides the purity consumed by the reverse step, with the removed entropy carried away by the traced-out register. Choosing $\tau_t$ pure maximizes the available budget in Eq.~(\ref{eq: prop1 budget}), so each step has the largest entropy-removal capacity allowed by the discarded register. This requirement follows from CPTP constraints on density operators rather than from a modeling preference. Proposition~2 extends this one-step budget in the pure-ancilla setting to an end-to-end constraint on diffusion depth and register width.

\subsection{An Entropy-Removal Budget for the Discarded Register}
\label{subsec: entropy budget}

Proposition~1 bounds the entropy that can be removed in one backward step. When specialized to the pure ancilla used by QGDM and accumulated over the full reverse trajectory, this one-step budget constrains the joint choice of $T$ and $N_{\tau}$.

\smallskip
\noindent\textbf{Proposition~2} (Depth--width budget for entropy removal).
\textit{Consider the denoising process $\rho_{t-1}=\operatorname{tr}_B\left(U(\tau_t\otimes\rho_t)U^{\dagger}\right)$, where $\tau_t$ is a pure state on $N_{\tau}$ qubits and $U$ is an arbitrary unitary. Then}
\[
S(\rho_t)-S(\rho_{t-1}) \le N_{\tau}\ln 2 .
\]
\textit{Consequently, if a sequence of $T$ such steps, with arbitrary and possibly different unitaries and pure embedding states, maps $\rho_T=\mathbb{I}/d$ to an $N$-qubit target $\rho_0$, then}
\begin{equation}
	\label{eq: trajectory entropy budget}
	TN_{\tau}\ge N-S(\rho_0)/\ln 2.
\end{equation}

\smallskip
\noindent\textit{Proof.}  A detailed proof is provided in the Supplementary Material. It specializes the one-step entropy budget in Proposition~1 to a pure ancilla, yielding $S(\rho_t)-S(\rho_{t-1}) \le N_\tau \ln 2$. Summing this bound over the $T$ reverse steps and using $S(\rho_T)=N\ln 2$ gives the constraint on $T$ and $N_\tau$.\hfill$\square$
\medskip

Eq.~(\ref{eq: trajectory entropy budget}) states that the total entropy gap $N-S(\rho_0)/\ln 2$ between the maximally mixed initialization and the target is bounded by the product of diffusion depth and discarded-register width. This constraint has three implications. First, $N_{\tau}=0$ gives a zero entropy-removal budget, consistent with Proposition~1. Second, a single-step generator ($T=1$) requires $N_{\tau}\ge N-S(\rho_0)/\ln 2$, which is close to a full-width register for near-pure targets. In contrast, a multi-step process distributes the same entropy removal across timesteps, allowing a smaller per-step register. Third, pure targets maximize $N-S(\rho_0)/\ln 2$ and are therefore the most demanding in this entropy-budget sense.

These are necessary architectural constraints rather than tight characterizations. For the default setting used in our experiments, $T=30$ and $N_{\tau}=N$, the budget is far from saturated, so the proposition should not be read as determining these hyperparameters. The same argument applies to any Stinespring-dilated architecture, including RQGDM, with $N_{\tau}$ replaced by the number of qubits discarded at each step.

\subsection{End-to-End Generation Guarantee of the Timestep-Wise Objective}
\label{subsec: e2e guarantee}

The training objective in Eqs.~(\ref{eq: loss function1})--(\ref{eq: loss function3}) penalizes each reverse step separately along the forward trajectory, whereas generation applies the learned steps sequentially to generated outputs. The result below shows that the per-step objective still controls the final generation error, with additive error accumulation. Let $D_{\mathrm{tr}}(\rho,\sigma)=\frac{1}{2}\lVert\rho-\sigma\rVert_1$ denote the trace distance.

\noindent\textbf{Proposition~3} (Per-step losses control the generation error). \textit{For $t=T,\dots,1$, let $f_{\boldsymbol{\Theta}}(\cdot,t)$ be the backward process at timestep $t$, $\rho_t$ be the forward states of Theorem~1, and let the generation iterates $\tilde{\rho}_t$ be initialized at $\tilde{\rho}_T=\mathbb{I}/d$ with $\tilde{\rho}_{t-1}=f_{\boldsymbol{\Theta}}(\tilde{\rho}_t,t)$. Under the cosine schedule $\bar{\alpha}_T=0$, the trace distance between the generated state $\tilde{\rho}_0$ and the target state $\rho_0$ satisfies:} 
\begin{equation} 
	\label{eq: end-to-end bound} D_{\mathrm{tr}}(\tilde{\rho}_0,\rho_0) \;\le\; \sum_{t=1}^{T}\sqrt{\mathcal{L}_{t-1}},
\end{equation} 
\textit{where $\mathcal{L}_{t-1}=1-F\!\left(\rho_{t-1},f_{\boldsymbol{\Theta}}(\rho_t,t)\right)$ are the per-timestep fidelity losses of Eq.\,(\ref{eq: loss function2}).}

\smallskip\noindent\textit{Proof.} A detailed proof is provided in the Supplementary Material. It uses the contractivity of trace distance under quantum channels, so a reverse step cannot increase an upstream error. A triangle-inequality recursion then telescopes the per-step errors into an additive bound, and the Fuchs--van de Graaf inequality converts these trace-distance terms into the per-timestep fidelity losses.\hfill$\square$
\medskip

Proposition~3 shows that the timestep-wise objective is consistent with end-to-end generation. If the per-step fidelity losses approach zero, the generated state approaches the target, and errors accumulate additively rather than multiplicatively. The key ingredient is that every reverse step is a CPTP map, for which trace distance is contractive. Thus, the learned physical channel cannot amplify errors inherited from earlier timesteps.

This bound also supports the use of the fidelity-based loss. The argument relies on trace-distance contractivity and its relation to fidelity through the Fuchs--van de Graaf inequality. In contrast, the Hilbert--Schmidt distance is not contractive under general CPTP maps~\cite{ozawa2000entanglement} and would not yield the same guarantee, consistent with its diagnostic role in Fig.~\ref{fig:nature design is bad}. Eq.~(\ref{eq: end-to-end bound}) should be read as a consistency guarantee rather than a tight numerical certificate; the realized generation fidelities are evaluated directly in Section~\ref{section: simulations}.

\subsection{Critical Evaluation of the Backward Process}

A straightforward design for the denoising process $f_{\boldsymbol{\Theta}}(\rho_t,t)$ involves using a trainable circuit $U(\boldsymbol{\theta})$ that acts on the extended system $\tau_t\otimes\rho_t$ and then traces out $\tau_t\text{-subsystem}$ ($i.e.$, all the qubits of $\tau_t$), which can be described as:
\begin{equation}
	\label{eq: bad design}
	\text{tr}_{\tau_t\text{-subsystem}}\left(U(\boldsymbol{\theta})\left(\tau_t \otimes \rho_t\right)U^{\dagger}(\boldsymbol{\theta})\right)=\tilde{\rho}_{t-1}.
\end{equation}
However, the noise added to the quantum state in a diffusion step is negligible, resulting in only a trivial difference between $\rho_t$ and $\rho_{t-1}$.
In the case of tracing out the register of $\tau_t$ and aiming for $\tilde{\rho}_{t-1} \approx \rho_t$, it can mislead the optimizer, which tends to adjust the circuit $U(\boldsymbol{\theta})$ to produce an output that closely approximates its input, thus minimizing the training loss. However, this adjustment deviates from the intended role of the denoising circuit.

\medskip
\noindent The following proposition characterizes the copy shortcut caused by the natural output-register choice.

\smallskip
\noindent\textbf{Proposition~4} (Low-loss copy shortcut in the natural design versus a strictly positive lower bound in the disjoint design).
\textit{Let $\rho_t$ be the QGDM forward chain of Eq.\,(\ref{eq: one step diffusion}) with $\bar{\alpha}_0=1$.}

\textit{(i) Natural design. In the natural output rule of Eq.\,(\ref{eq: bad design}), every product unitary $U=U_{\tau}\otimes I_{\rho}$ yields the exact-copy output $\tilde{\rho}_{t-1}=\rho_t$. For the QGDM objective in Eq.\,(\ref{eq: loss function1}), this copy shortcut satisfies
\begin{equation}
	\mathcal{L}
	\le
	2(1-2^{-N})
	\left[
	(1-\bar{\alpha}_1)
	+
	\frac{\lambda\bar{\alpha}_1}{T-1}
	\right]
	\le
	2(1-\bar{\alpha}_1)
	+
	\frac{2\lambda}{T-1}.
	\label{eq:dimension-uniform-copy-bound}
\end{equation}
If this copy map is used during generation from $\rho_T=\mathbb{I}/d$, every reverse iterate remains $\mathbb{I}/d$.}

\textit{(ii) Disjoint-register design. When the output is read from the timestep register, every product unitary $U=U_{\tau}\otimes U_{\rho}$ yields the input-independent pure state $\tilde{\rho}_{t-1}=U_{\tau}\tau_tU_{\tau}^{\dagger}$. Consequently, for any mixed target $\rho_0$ the entire product-unitary family obeys the strictly positive lower bound
\begin{equation}
	\mathcal{L}\;\ge\;1-\lambda_{\max}(\rho_0)>0,
	\label{eq:disjoint-loss-floor}
\end{equation}
where $\lambda_{\max}(\rho_0)$ denotes the largest eigenvalue of $\rho_0$ and is unrelated to the loss hyperparameter $\lambda$.}

\smallskip
\noindent\textit{Proof.}\;A detailed proof is provided in the Supplementary Material. It compares the same product-unitary family under the two register layouts. In the natural layout, this family copies the noisy input, whose proximity to $\rho_{t-1}$ along the forward chain gives a small objective value. In the disjoint layout, the same product-family structure produces an input-independent pure state, whose overlap with a mixed target is bounded by $\lambda_{\max}(\rho_0)$, giving the strictly positive lower bound.\hfill$\square$

\medskip
Proposition~4 compares the same product-unitary family under two output-register designs. In the natural design, every $U=U_{\tau}\otimes I_{\rho}$ exactly copies the noisy input. Because adjacent forward states are close, this copy family can attain the small objective value in Eq.~(\ref{eq:dimension-uniform-copy-bound}). However, it fails during generation: if the process is initialized at $\rho_T\approx\mathbb{I}/d$, copying keeps all reverse iterates close to the maximally mixed state. In the disjoint design, every $U=U_{\tau}\otimes U_{\rho}$ instead outputs a pure state independent of $\rho_t$. For a mixed target, this product-unitary family is bounded below by Eq.~(\ref{eq:disjoint-loss-floor}), so it cannot become the same low-loss copy shortcut.

This result does not rule out all copy-like maps. For example, a SWAP-like unitary can still move $\rho_t$ into the output register. The point is more specific: the disjoint layout removes the simplest passive product-unitary shortcut that is available in the natural layout. In the natural design, copying is already a low-loss solution within the product family. In the disjoint design of Eq.~(\ref{eq:denoising process}), the product family is penalized on mixed targets, so reducing the loss requires cross-register information transfer beyond the $U_{\tau}\otimes U_{\rho}$ form. This explains why the output-register choice is important for avoiding the observed training failure.

\begin{figure}[h]
	\centering
	\includegraphics[width=0.9\linewidth]{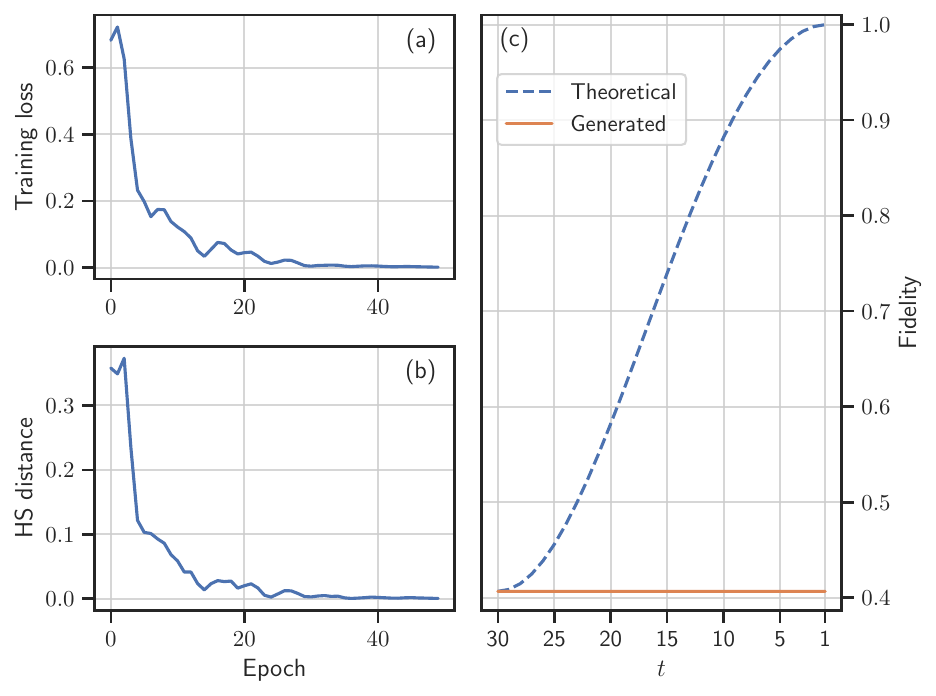}
	\caption{(a) Training loss with respect to the iterations of training. (b) Hilbert--Schmidt (HS) distance between $\rho_t$ and $\tilde{\rho}_{t-1}$ in the failed denoising design. It is used only to diagnose input-copying behavior, not as the training loss of QGDM. (c) After training, the generation fidelity of QGDM with respect to the timestep.}
	\label{fig:nature design is bad}
\end{figure}

The numerical simulation results illustrating the limitations of natural design (Eq.~(\ref{eq: bad design})) are shown in Fig.~\ref{fig:nature design is bad}. Fig.~\ref{fig:nature design is bad}(a) displays the decreasing trend of the training loss over epochs. 
Fig.~\ref{fig:nature design is bad}(b) reports the Hilbert--Schmidt (HS) distance $D_{\mathrm{HS}}(\rho_t,\tilde{\rho}_{t-1})=\sqrt{\mathrm{tr}\left[(\rho_t-\tilde{\rho}_{t-1})^2\right]}$ between the noisy input state $\rho_t$ and the circuit output $\tilde{\rho}_{t-1}$ during training. This HS distance serves only as a diagnostic quantity for this failed design and is not the training objective of QGDM. The training objective remains the fidelity-based loss in Eqs.\,(\ref{eq: loss function1})--(\ref{eq: loss function3}). The HS distance is plotted to check, at the density-matrix level, whether the circuit output simply copies its input. Fig.~\ref{fig:nature design is bad}(c) further shows the generation fidelity of QGDM with respect to timestep $t$. The orange solid line represents the fidelity of the generated states with the target state, while the blue dashed line represents the theoretical fidelity curve that the model should achieve.

Figs.\,\ref{fig:nature design is bad}(a)-(b) indicate that with increasing epochs, both training loss and HS distance rapidly decrease. A reduction in this distance indicates that the reduced circuit output $\tilde{\rho}_{t-1}$ approaches the noisy input state $\rho_t$, ultimately becoming nearly identical to it. Fig.\,\ref{fig:nature design is bad}(c) shows that QGDM cannot alter the input state during the generation process. The denoising process merely outputs the quantum state fed into it. Hence the fidelity between the generated quantum state and target remains unchanged.

\begin{figure}[h]
	\centering
	\includegraphics[width=0.9\linewidth]{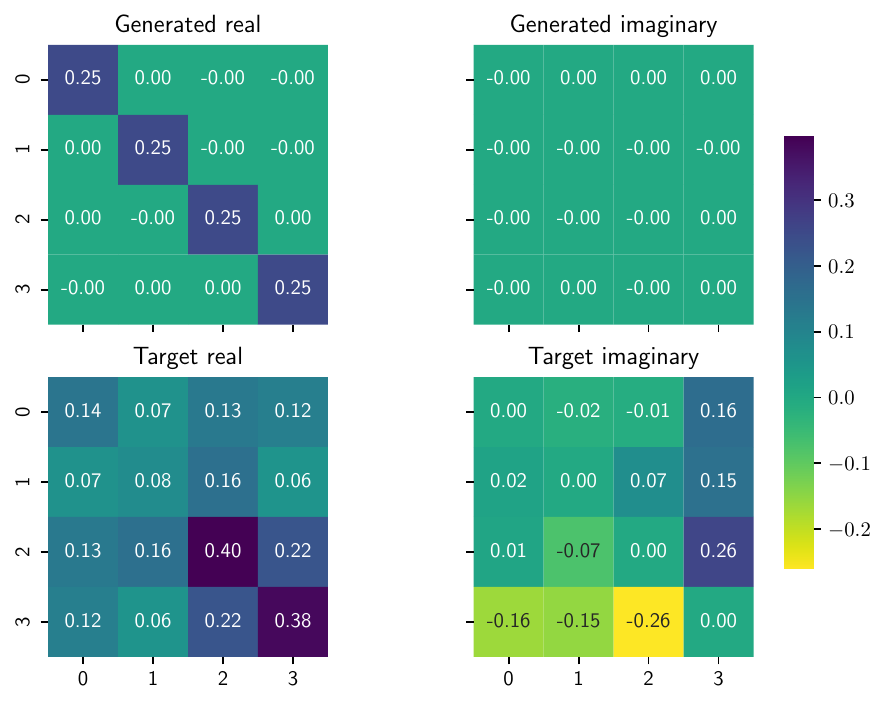}
	\caption{Visualization of the generated density matrix and target density matrix.}
	\label{fig:bad generation}
\end{figure}

The comparison between the final generated quantum state $\tilde{\rho}_0$ (shown in the first row) and $\rho_0$ (displayed in the second row) is illustrated in Fig.\,\ref{fig:bad generation}. Given the complex nature of the density matrix, the figure separately draws the real and imaginary parts, with the former on the left and the latter on the right. This visualization clearly shows that $\tilde{\rho}_0$ remains equal to the completely mixed state $\rho_T$, indicating that $U(\boldsymbol{\theta})$ essentially reproduces the input state. This highlights the critical need for a well-designed denoising process in QGDM that effectively integrates information between the noisy input state $\rho_t$ and the timestep embedding state $\tau_t$, especially considering the minor differences in quantum states across adjacent timesteps. The results and discussions in our study intend to provide valuable insights for the future development of similar quantum machine learning models.

Finally, we observe that $U(\boldsymbol{\theta})$ is not necessarily optimized to identity matrix $\mathbb{I}$ every time, although it seems straightforward to use $\mathbb{I}$ as a simple way to keep input and output states the same. Learning algorithms do not always choose to do so. 
This is because different ensembles of quantum states can be transformed into each other by $U(\boldsymbol{\theta})$ while maintaining the same density matrix. There may be more than one transform in the hypothesis space that can satisfy this condition. After training, which one of them is chosen depends on the position of $U(\boldsymbol{\theta})$ in the hypothesis space when the parameters $\boldsymbol{\theta}$ are initialized.

\section{Numerical Simulations}
\label{section: simulations}

Our numerical experiments cover two complementary benchmarks. The first uses random-circuit-generated pure and mixed states to evaluate generative performance without structural priors, following the standard use of random quantum objects as probes in quantum information~\cite{emerson2003pseudo,harrow2009random,brandao2016local,bouland2019complexity}. On this benchmark, we compare our models with QuGAN~\cite{dallaire2018quantum} and EQ-GAN~\cite{niu2022entangling}.

The second benchmark uses Gibbs states of the transverse-field Ising model (TFIM) as physically structured mixed-state targets. It tests QGDM on a thermal-state generation task and compares it with stronger baselines, including VGSP~\cite{consiglio2024variational}, QuDT, and $\zeta$-QVAE~\cite{wang2025quantum}.

For more details on the numerical simulations, including training hyperparameters, target state preparation, and the training and generation metrics of our models, please refer to Supplementary File. 

\subsection{Setup}

We use the Tensorcircuit framework \cite{zhang2023tensorcircuit} to simulate quantum circuits and the Tensorflow framework \cite{tensorflow2015_whitepaper} for parameter optimization, with the optimizer Adam \cite{kingma2014adam}. 
During training, a larger learning rate may make it difficult for the model to converge, while setting the learning rate too small makes the model converge too slowly. To better optimize the model, we use a cosine decay method \cite{loshchilov2016sgdr} to dynamically adjust the learning rate.

Our numerical simulations cover a number of qubits ranging from 1 to 8.
QGDM requires $2N$ qubits to generate an $N$-qubit target state. Therefore, in our simulations, QGDM achieves results for $N \le 4$.
EQ-GAN can generate mixed states for $N \le 7$, as simulating the $N=8$ task demands an unattainable amount of computer memory.
At $N=1$, RQGDM degenerates into QGDM.
Therefore, in the next two subsections, the results for QGDM at $N>4$, RQGDM at $N=1$, and EQ-GAN at $N=8$ in mixed-state generation are excluded. 
All simulations are repeated 10 times with different random seeds to obtain the statistical characteristics of our results.

\subsection{Random State Generation}

\subsubsection{Impact of $N_{\tau}$}
\begin{figure}[ht]
	\centering
	\includegraphics[width=0.9\linewidth]{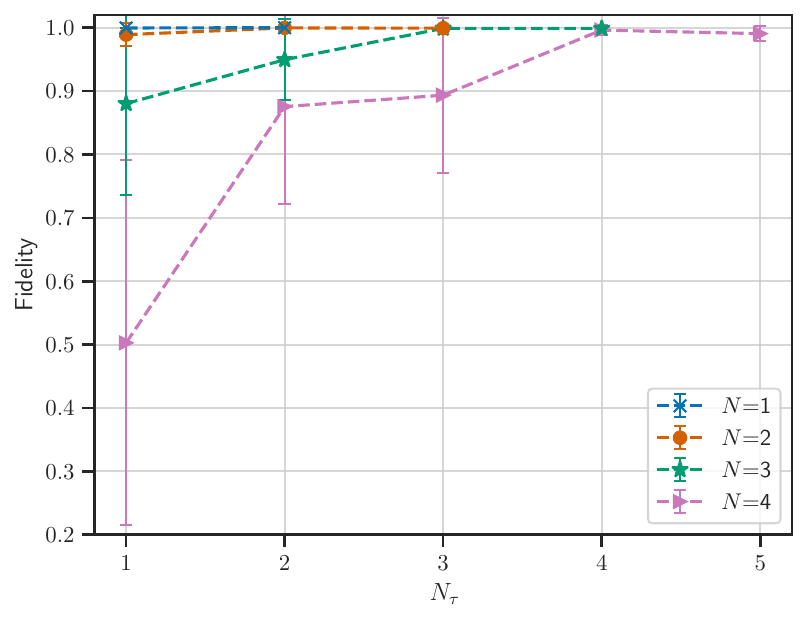}
	\caption{The generation fidelity of QGDM in generating target pure states for different qubits with respect to $N_{\tau}$.}
	\label{fig:relation_Fidelity_NTime}
\end{figure}

We first investigate the impact of $N_{\tau}$ in QGDM for generating pure states. The results are shown in Fig.\,\ref{fig:relation_Fidelity_NTime}.
Generation fidelity improves with $N_{\tau}$ when $N_{\tau} \le N$, achieving the maximum when $N_{\tau} = N$ for all $N$. 
\begin{table*}[t]  
	\caption{Comparison of pure state generation between QuGAN, EQ-GAN, and our proposed models.}  
	\label{tab: pure state}  
	\centering  
	\begin{tabular}{|c|c|c|c|c|c|c|c|c|c}
		\hline  
		Models & $N=1$ & $N=2$ & $N=3$ & $N=4$ & $N=5$ & $N=6$ & $N=7$ & $N=8$\\  
		\hline  
		QuGAN \cite{dallaire2018quantum} & 0.999$\pm$5e-5 & 0.996$\pm$4e-3 & 0.995$\pm$4e-3 & 0.993$\pm$3e-3 & 0.990$\pm$4e-3 & 0.986$\pm$3e-3 & 0.978$\pm$6e-3 & 0.971$\pm$1e-1\\
		\hline  
		EQ-GAN \cite{niu2022entangling} & \textbf{1.00$\pm$0} & 0.978$\pm$5e-2 & 0.947$\pm$4e-2 & 0.896$\pm$8e-2 & 0.898$\pm$9e-2 & 0.872$\pm$9e-2 & 0.833$\pm$1e-1 & 0.824$\pm$8e-2\\
		\hline
		QGDM (Ours) & 0.999$\pm$3e-5 & \textbf{0.999$\pm$2e-4} & \textbf{0.999$\pm$2e-3} & 0.996$\pm$8e-3 & / & / & / & / \\
		\hline  
		RQGDM (Ours) & / & 0.994$\pm$8e-3 & 0.999$\pm$6e-4 & \textbf{0.999$\pm$4e-4} & \textbf{0.997$\pm$4e-3} & \textbf{0.995$\pm$1e-2} & \textbf{0.994$\pm$4e-3} &
		\textbf{0.990$\pm$1e-2} \\
		\hline  
	\end{tabular}  
\end{table*}

However, when $N_{\tau}>N$, except for the case $N=1$, fidelity decreases slightly in other cases. The reason for this may be that there is a redundant qubit that complicates the denoising process. 
The denoising circuit aims to predict the denoising state $\tilde{\rho}_t$ by merging information from $\tau_t$ and $\rho_t$, and then projecting its prediction onto the first $N$ qubits of the $\tau_t$ state. These qubits are denoted as register \textit{A} in Fig.\,\ref{fig:original}.
In cases where $N_{\tau} > N$, there is an extra qubit in the state $\tau_t$ compared to $\rho_t$. This additional qubit may be redundant, complicating the process of merging and output prediction. 

When $N_{\tau} < N$, register \textit{A} contains a part of the qubits of $\rho_t$. 
However, the presence of overlapping qubits between register \textit{A} and the qubits of $\rho_t$ may diminish the efficacy of $U(\boldsymbol{\theta})$ in such scenarios since a part of the information from $\rho_t$ is already stored in \textit{A}. In Section~\ref{section: design about f}, we have investigated a well-intentioned but faulty denoising design. This design involves transferring all qubits from $\rho_t$ to the output register, maximizing the overlap between \textit{A} and the qubits of $\rho_t$. This extreme case strengthens our hypothesis, illustrating the potential implications.

Based on the results in Fig.\,\ref{fig:relation_Fidelity_NTime}, subsequent simulations default to $N_{\tau}=N$ as this setting yields the best generative results with our proposed method.
This choice is also consistent with Proposition~4(ii): when $N_\tau = N$, the $N$-qubit output register $A$ is fully disjoint from the $N$-qubit noisy-input register carrying $\rho_t$. The empirical results therefore support the architectural rationale that this disjoint-register design suppresses the simplest passive copy shortcut while retaining the capacity for input-dependent denoising.

\subsubsection{Pure States Generation Results}

\begin{figure}[h]
	\centering
	\includegraphics[width=0.9\linewidth]{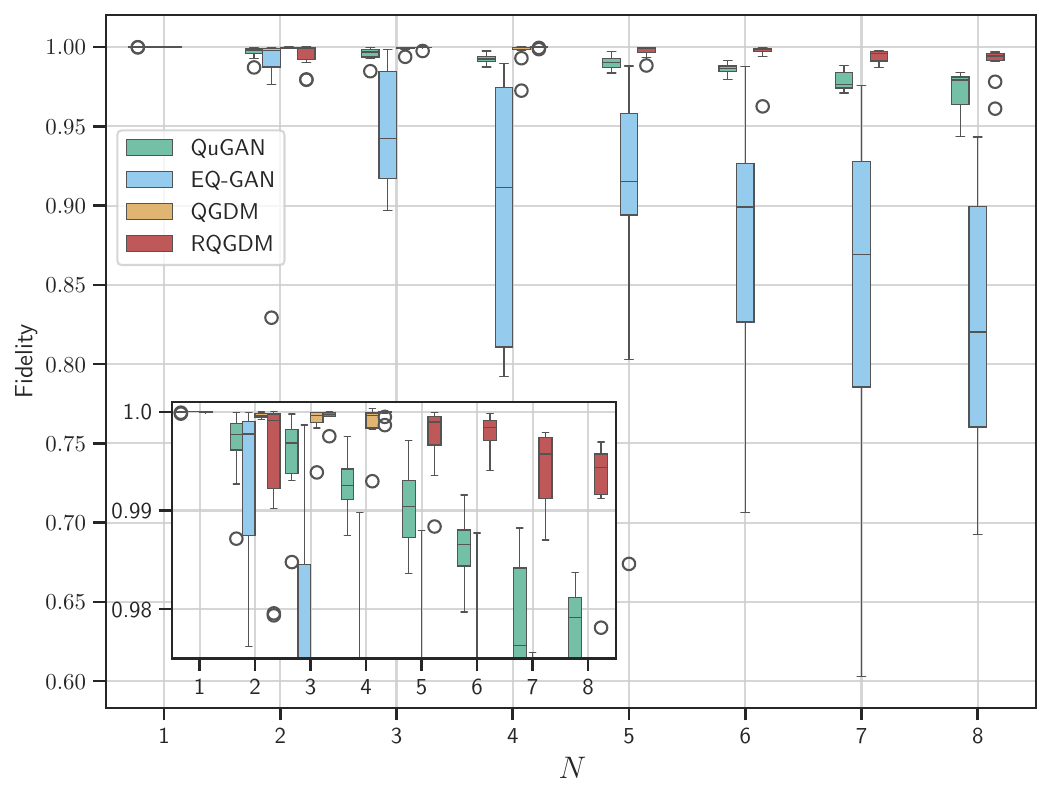}
	\caption{Pure state generation results. The circles represent outliers, which are data points that significantly deviate from the rest of the data.}
	\label{fig:ours qgan pure state}
\end{figure}

In Fig.\,\ref{fig:ours qgan pure state}, we compare QGDM and RQGDM with QuGAN \cite{dallaire2018quantum} and EQ-GAN \cite{niu2022entangling} to assess their fidelity in generating the desired pure state. The results indicate that as the number of qubits in the target state increases, the fidelity of all models decreases.

QuGAN maintains high fidelity at lower qubit counts. However, its median fidelity decreases as the number of qubits increases, along with an increase in variance. For instances with $N\ge7$, QuGAN's fidelity falls below 0.99 for all 10 experiments, indicating its limitations in handling more complex quantum states. EQ-GAN achieves a perfect fidelity of 1.0 when the system has only one qubit. However, its fidelity decreases as the number of qubits grows, displaying the largest variance among the four models. From 3 qubits onward, EQ-GAN's fidelity consistently stays below 0.99, highlighting its challenges in generating pure states. The fidelity of QGDM and RQGDM remains mostly above 0.99 across the tested qubit numbers, with more concentrated data distributions. In most scenarios, RQGDM surpasses the other three models in terms of median fidelity and demonstrates greater stability at higher tested qubit counts than its peers.
In summary, QGDM performs well in smaller quantum systems, while RQGDM provides a more resource-conscious alternative across the tested qubit configurations.

To better evaluate the performance differences among models, we present the generative fidelity values for each model in Table.\,\ref{tab: pure state}. The fidelity of QuGAN diminishes as the number of qubits increases, starting from 0.999$\pm$5e-5 at $N=1$ to 0.971$\pm$1e-1 at $N=8$. It is worth noting that its fidelity drops below 0.99 for $N \ge 6$, indicating a decline in performance as the quantum system size increases. The fidelity of QuGAN from $N=1$ to $5$ remains above 0.99, showing a downward trend for larger systems, although these values are consistently lower than those achieved by our proposed method.
EQ-GAN achieves the best initial result, producing fidelities of 1 at $N=1$. However, as the number of qubits increases, EQ-GAN shows the lowest average fidelity among the four models, along with the highest variance.
QGDM exhibits stable fidelity, with an average of 0.999 at $N=1, 2$, and $3$, and 0.996$\pm$8e-3 at $N=4$. Importantly, RQGDM consistently demonstrates high fidelity across all tested numbers of qubits, even achieving 0.99 at $N=8$. This indicates stable high-fidelity pure-state generation on the tested range.

\subsubsection{Timestep Embedding State Visualization}
\label{section: Timestep Embedding Visualization}

\begin{figure}[h]
	\centering
	\includegraphics[width=0.9\linewidth]{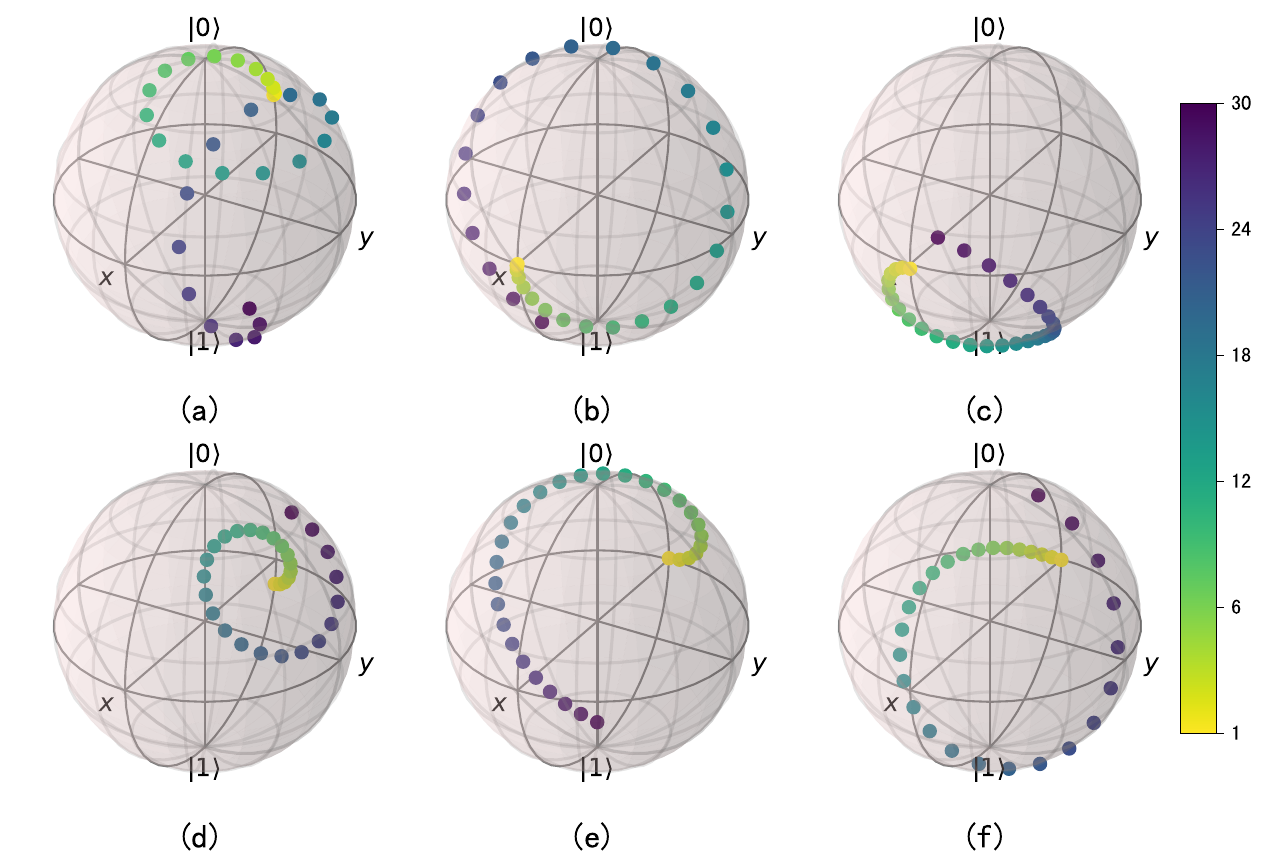}
	\caption{Selected examples visualizing single-qubit timestep embedding states.
		Our method allows the timestep embedding state to adaptively distribute itself on the surface of the Bloch sphere while maintaining temporal coherence. This demonstrates that our approach to encoding temporal information is expressive and flexible.}
	\label{fig:bloch spheres}
\end{figure}

Fig.\,\ref{fig:bloch spheres} shows selected examples visualizing a single-qubit timestep embedding state sequence $\{\tau_t\}_{t=1}^{T}$ after training, with $T=30$. Different colors are used to denote the states in each timestep $t$. Each example demonstrates that $\tau_t$ shifts its position on the Bloch sphere surface coherently as $t$ progresses. The density of the states varies as well.
Figs.\,\ref{fig:bloch spheres}(a)-(b) show a more dispersed distribution on the Bloch sphere surface. Figs.\,\ref{fig:bloch spheres}(c)-(d) illustrate cases where states are more clustered. Figs.\,\ref{fig:bloch spheres}(e)-(f) display an intermediate situation, with the state sequence moderately and coherently distributed over the surface of the Bloch sphere. 

Compared to qubit encoding methods \cite{stoudenmire2016supervised, grant2018hierarchical}, our approach to timestep embedding, as inspired by \cite{lloyd2020quantum}, is more expressive and flexible. 
The qubit encoding method generates states that are equidistantly arranged in a ring on the Bloch sphere surface. This restricts subsequent learning models to only learning from such fixed patterns. In contrast, our approach allows the timestep embedding states to be adaptable, enhancing their multifunctionality. This could potentially decrease the difficulty of learning for subsequent models and improve the overall model's performance.

\subsubsection{Mixed States Generation Results}

\begin{figure}[h]
	\centering
	\includegraphics[width=0.9\linewidth]{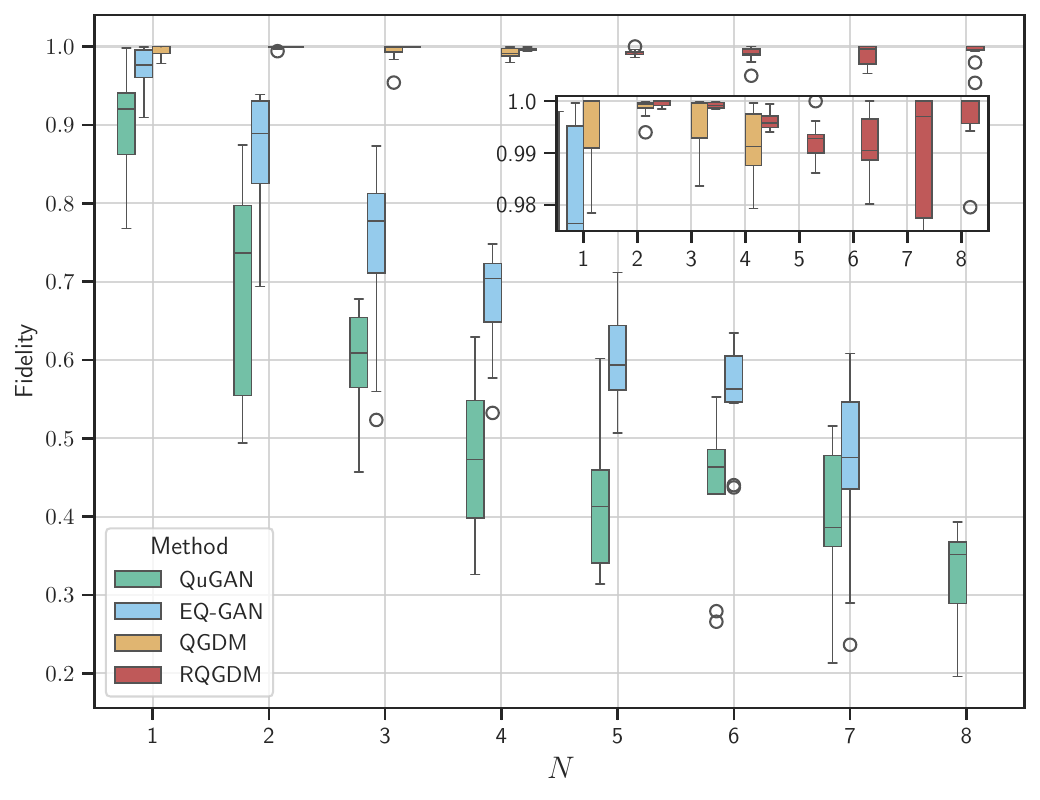}
	\caption{Mixed state generation results. The circles represent outliers, which are data points that significantly deviate from the rest of the data.}
	\label{fig:Ours qgan mixed state}
\end{figure}

Fig.\,\ref{fig:Ours qgan mixed state} and Table\,\ref{tab: Ours qgan mixed state} report the random mixed-state benchmark. These targets are low-rank mixtures of random-circuit-generated pure states, so this experiment should be read as a structure-free expressivity test rather than as evidence of performance over the full mixed-state manifold. Within this diagnostic setting, QuGAN and EQ-GAN degrade as $N$ increases, while QGDM and RQGDM maintain high fidelity in the tested cases where they are applicable. This diagnostic comparison illustrates the difficulty of adversarial quantum generative training on mixed-state targets, while the structured TFIM Gibbs benchmark below evaluates physically meaningful thermal targets.

\begin{table*}[t]  
	\caption{Comparison of mixed state generation performance between QuGAN, EQ-GAN, and our proposed models.}  
	\label{tab: Ours qgan mixed state}  
	\centering  
	\begin{tabular}{|c|c|c|c|c|c|c|c|c|}
		\hline  
		Models & $N=1$ & $N=2$ & $N=3$ & $N=4$ & $N=5$ & $N=6$ & $N=7$ & $N=8$\\ 
		\hline  
		QuGAN \cite{dallaire2018quantum} & 0.902$\pm$6e-2 & 0.694$\pm$1e-1 & 0.598$\pm$6e-2 & 0.477$\pm$9e-2 & 0.414$\pm$8e-2& 0.439$\pm$9e-2& 0.396$\pm$8e-2 & 0.324$\pm$6e-2 \\
		\hline  
		EQ-GAN \cite{niu2022entangling} & 0.972$\pm$3e-2 & 0.867$\pm$8e-2 & 0.741$\pm$1e-1 & 0.676$\pm$7e-2 & 0.602$\pm$6e-2 & 0.555$\pm$6e-2 & 0.463$\pm$1e-1 & /\\
		\hline
		QGDM (Ours) & \textbf{0.995$\pm$7e-3} & \textbf{0.999$\pm$1e-3} & 0.992$\pm$1e-2 & 0.992$\pm$7e-3 & / & / & / & / \\
		\hline  
		RQGDM (Ours) & / & 0.999$\pm$6e-4 & \textbf{0.999$\pm$5e-4} & \textbf{0.996$\pm$2e-3} & \textbf{0.993$\pm$4e-3} & \textbf{0.990$\pm$1e-2} & \textbf{0.992$\pm$2e-2} & \textbf{0.993$\pm$2e-2} 
		\\  
		\hline  
	\end{tabular}  
\end{table*}

The detailed values in Table\,\ref{tab: Ours qgan mixed state} support the same diagnostic conclusion. QuGAN starts at a fidelity of 0.902$\pm$6e-2 at $N=1$ and declines to 0.324$\pm$6e-2 at $N=8$. EQ-GAN generally performs better than QuGAN but also decreases with system size and is not reported for $N=8$ because of memory limits. QGDM is evaluated up to $N=4$ because it requires $2N$ qubits, while RQGDM extends the tested random mixed-state setting to $N=8$. These results motivate a non-adversarial mixed-state generator, but we avoid using this benchmark alone as the basis for broad superiority claims.


\begin{table*}[t]
	\caption{Aggregate TFIM Gibbs-state fidelity under gate-level depolarizing noise. Results are grouped by two-qubit gate-noise strength $\eta\in\{0,0.5\%,1\%\}$ and target size $N$, with QGDM placed at the bottom for ease of comparison. Each entry reports mean$\pm$standard deviation over $h\in\{0.5,1.0,1.5\}$, $\beta\in\{0.5,1,2,3,4,5\}$, and ten random initialization seeds. Single-qubit gate noise is set to $\eta/10$.	}
	\label{tab:gibbs_fidelity_main}
	\centering
	\scriptsize
	\setlength{\tabcolsep}{1.6pt}
	\renewcommand{\arraystretch}{1.03}
	\begin{tabular}{@{}|l|cccc|cccc|cccc@{}|}
		\hline
		& \multicolumn{4}{@{}c|@{}}{\textbf{$\eta=0$}} & \multicolumn{4}{@{}c|@{}}{\textbf{$\eta=0.5\%$}} & \multicolumn{4}{@{}c|@{}}{\textbf{$\eta=1\%$}}\\
		\cline{2-13}
		Model & $N=1$ & $N=2$ & $N=3$ & $N=4$ & $N=1$ & $N=2$ & $N=3$ & $N=4$ & $N=1$ & $N=2$ & $N=3$ & $N=4$\\
		\hline
		QuGAN~\cite{dallaire2018quantum} & 0.726$\pm$1e-1 & 0.478$\pm$2e-1 & 0.334$\pm$2e-1 & 0.262$\pm$2e-1 & 0.726$\pm$1e-1 & 0.485$\pm$2e-1 & 0.355$\pm$2e-1 & 0.291$\pm$2e-1 & 0.726$\pm$1e-1 & 0.487$\pm$2e-1 & 0.366$\pm$2e-1 & 0.298$\pm$2e-1\\
		\hline
		EQ-GAN~\cite{niu2022entangling} & 0.666$\pm$2e-1 & 0.444$\pm$2e-1 & 0.295$\pm$2e-1 & 0.263$\pm$1e-1 & 0.667$\pm$2e-1 & 0.457$\pm$2e-1 & 0.322$\pm$2e-1 & 0.286$\pm$2e-1 & 0.667$\pm$2e-1 & 0.467$\pm$2e-1 & 0.332$\pm$2e-1 & 0.290$\pm$2e-1\\
		\hline
		$ \zeta $-QVAE~\cite{wang2025quantum} & 0.908$\pm$5e-2 & 0.894$\pm$5e-2 & 0.762$\pm$1e-1 & 0.669$\pm$2e-1 & 0.905$\pm$5e-2 & 0.885$\pm$5e-2 & 0.743$\pm$1e-1 & 0.650$\pm$2e-1 & 0.903$\pm$6e-2 & 0.876$\pm$6e-2 & 0.722$\pm$1e-1 & 0.625$\pm$2e-1\\
		\hline
		QuDT & 0.998$\pm$3e-3 & 0.909$\pm$1e-1 & 0.829$\pm$2e-1 & 0.731$\pm$2e-1 & 0.997$\pm$3e-3 & 0.903$\pm$1e-1 & 0.815$\pm$2e-1 & 0.710$\pm$2e-1 & \textbf{0.996$\pm$4e-3} & 0.898$\pm$1e-1 & 0.801$\pm$2e-1 & 0.697$\pm$2e-1\\
		\hline
		VGSP~\cite{consiglio2024variational} & \textbf{1.000$\pm$0} & \textbf{0.999$\pm$1e-4} & \textbf{0.999$\pm$1e-3} & \textbf{0.999$\pm$3e-3} & 0.998$\pm$3e-3 & 0.971$\pm$2e-2 & 0.944$\pm$4e-2 & 0.905$\pm$6e-2 & 0.995$\pm$6e-3 & 0.939$\pm$4e-2 & 0.886$\pm$7e-2 & 0.813$\pm$1e-1\\
		\hline
		QGDM (Ours) & 0.999$\pm$4e-3 & 0.998$\pm$4e-3 & 0.994$\pm$1e-2 & 0.991$\pm$1e-2 & \textbf{0.998$\pm$3e-3} & \textbf{0.980$\pm$1e-2} & \textbf{0.956$\pm$3e-2} & \textbf{0.913$\pm$5e-2} & 0.995$\pm$6e-3 & \textbf{0.960$\pm$3e-2} & \textbf{0.915$\pm$6e-2} & \textbf{0.837$\pm$1e-1}\\
		\hline
		
	\end{tabular}
\end{table*}

\subsection{Structured Gibbs-State Generation}
To complement the random-state benchmarks, we evaluate QGDM on Gibbs states of the transverse-field Ising model (TFIM),
\begin{equation}
	\begin{aligned}
		\rho_{\beta}&=\exp(-\beta H)/\operatorname{tr}[\exp(-\beta H)],\\
		H&=-J\sum_i X_iX_{i+1}-h\sum_i Z_i,
	\end{aligned}
\end{equation}
which provide structured mixed-state targets relevant to thermal-state preparation and noisy quantum simulation. We benchmark $N=1,\ldots,4$ target qubits across $h\in\{0.5,1.0,1.5\}$ and $\beta\in\{0.5,1,2,3,4,5\}$. Gate-level noise is modeled by two-qubit depolarizing strengths $\eta\in\{0,0.5\%,1\%\}$, with single-qubit gate noise set to $\eta/10$, consistent with noise levels reported for recent superconducting devices~\cite{jin2025topological}.

We compare with baselines from four methodological families: QuDT (QCBM-based), $\zeta$-QVAE~\cite{wang2025quantum} (quantum VAE-based), QuGAN~\cite{dallaire2018quantum} and EQ-GAN~\cite{niu2022entangling} (QGAN-based), and the variational Gibbs state preparation method of Consiglio et al.~\cite{consiglio2024variational}, denoted as VGSP for brevity.
QuDT is a diffusion-free baseline constructed following~\cite{zhang2024generative}: it directly prepares a target mixed state through a parameterized joint circuit and partial trace. It shares the same ansatz family as QGDM but removes the multi-step diffusion-denoising process, and therefore serves as an ablation of that process. RQGDM is excluded because its compressed-register design is intended for low-rank mixed states; full-rank Gibbs states would require enlarging the compressed register, eliminating the intended auxiliary-qubit saving.

Table~\ref{tab:gibbs_fidelity_main} reports aggregate fidelity averaged over $h$, $\beta$, system size $N$, and random seeds. In noiseless simulation, the Hamiltonian-specialized VGSP achieves near-perfect fidelity ($0.999$ across the grid), and QGDM remains close at $0.996$. Under gate-level noise, QGDM attains the highest mean fidelity after averaging over $N$: $0.962$ at $\eta=0.5\%$ and $0.927$ at $\eta=1\%$, compared with VGSP's $0.954$ and $0.908$. The only per-$N$ exception is the single-qubit case at $\eta=1\%$, where QuDT ($0.996$) is marginally higher than QGDM ($0.995$). QuDT and $\zeta$-QVAE remain competitive at small $N$ but degrade more noticeably as $N$ increases, with $N$-averaged fidelities dropping to $0.867/0.856/0.848$ (QuDT) and $0.808/0.796/0.782$ ($\zeta$-QVAE) across the three noise levels.

The same conclusion is obtained from a paired block-level view over the eight nonzero-noise $(N,\eta)$ aggregate blocks in Table~\ref{tab:gibbs_fidelity_main}. QGDM is the top method in seven of these eight blocks; the only exception is $N=1$ at $\eta=1\%$, where QuDT is higher by $0.001$. Relative to VGSP, QGDM's averaged fidelity gain is $+0.0076$ at $\eta=0.5\%$ and $+0.0186$ at $\eta=1\%$, with positive gains for all $N\ge2$ at both nonzero noise levels. This paired view complements the reported standard deviations and shows that the noisy-regime advantage is not produced by a single favorable system size.

The QGAN-based baselines lag behind on this thermal-state task. Their slight fidelity increase under stronger noise appears to be an artifact of depolarizing gate noise, which mixes the outputs toward the maximally mixed state and can incidentally increase their overlap with mixed Gibbs targets. The diffusion-free QuDT baseline achieves consistently lower fidelity than QGDM across all tested noise levels, supporting the contribution of the multi-step diffusion-denoising process beyond the shared variational ansatz. All methods are evaluated on the same Hamiltonian and noise grids using the same fidelity metric and seed count; per-method circuit settings and results resolved over individual $h$ and $\beta$ values are provided in Supplementary Material.

\section{Conclusion}
\label{section: conclusion}

This paper introduces QGDM, a fully quantum-mechanical diffusion model for generating quantum states across a broad range of mixedness. Its fundamental concept is that any target quantum state can be converted into a completely mixed state through a non-unitary forward process. Following this, a trainable backward process is employed to reconstruct the target state from the completely mixed state. The backward process is realized as a variational quantum channel built from an auxiliary register, a joint unitary, and a partial trace, so that every denoising step is physically valid and capable of producing mixed-state outputs. The denoising process involves a parameterized function composed of a timestep embedding circuit and a denoising circuit, employing the concept of parameter sharing. Specifically, the timestep embedding circuit embeds temporal information into a quantum state, while the denoising circuit processes the polluted quantum state together with the timestep embedding state to produce a denoised state. To minimize auxiliary qubits in QGDM, we introduce a resource-efficient version of QGDM called RQGDM.

These design choices follow from the structure of quantum channels rather than analogy with classical diffusion alone. Since the forward depolarizing step has no exact CPTP inverse, the backward process cannot be obtained by direct inversion; instead, QGDM learns an approximate physical channel on the forward-generated states. Our analysis shows that this channel is subject to several structural constraints. Reducing mixedness requires drawing purity from a non-maximally mixed ancilla; the entropy removable per step is bounded by the diffusion depth and register width, motivating the multi-step process over single-step generation; and reading the output from a register disjoint from the noisy input removes a direct copy shortcut that the natural layout would otherwise permit. Since each reverse step is a valid quantum channel, timestep-wise fidelity losses compose without amplification, giving an end-to-end generation guarantee.

We evaluate QGDM on two complementary simulator-based benchmarks. On random-circuit-generated pure and mixed targets serving as structure-free expressivity tests, QGDM and RQGDM achieve high fidelity across the tested system sizes. On structured Gibbs states of the transverse-field Ising model under gate-level depolarizing noise, QGDM, as a general-purpose mixed-state generator, is competitive with the Hamiltonian-specialized VGSP in noiseless simulation and achieves the highest mean fidelity among the tested methods at nonzero noise levels. This noisy-regime behavior is also stable at the block level: QGDM is the top method in seven of the eight nonzero-noise $(N,\eta)$ aggregate blocks. The consistent advantage over the diffusion-free QuDT baseline, which shares the same variational ansatz but omits the multi-step process, further supports the interpretation that the diffusion mechanism contributes to the performance gain rather than the ansatz alone. These results identify a channel-based route to robust mixed-state generation in the tested simulator regime, with the Gibbs benchmark reaching four target qubits and the random-state benchmarks reaching eight.

Future work should explore the impact of different noise schedules on generation performance. Additionally, designing more efficient structures for the denoising process to reduce resource requirements, including the number of auxiliary qubits and the total number of quantum gates, is a worthwhile direction to pursue. Investigating the relationship between problem properties and the number of different time steps can provide important insights for further understanding QGDM. A deeper analysis of approximation error, sample complexity, and optimization landscapes for diffusion-style quantum generators would further strengthen the theoretical foundation. Validation under more comprehensive hardware noise models, including coherent errors, crosstalk, and readout noise, is an important step toward deployment on physical devices. More broadly, since mixed density operators are the native description of thermalized systems, open-system dynamics, and noisy near-term devices, QGDM provides a physically grounded and trainable route to generating such states through valid channel dynamics.

\ifCLASSOPTIONcompsoc
  \section*{Acknowledgments}
\else
  \section*{Acknowledgment}
\fi

This work was supported by the Science and Technology Development Fund, Macau SAR under Grant 0077/2025/RIA2, Grant 0038/2026/RIB1, and Grant 0007/2026/ASJ, National Natural Science Foundation of China (62471187), Guangdong Basic and Applied Basic Research Foundation (Grant 2025A1515011489), and the Guangdong Provincial Quantum Science Strategic Initiative (Grants GDZX2503001).
The authors are thankful to Prof. Wenmin Wang for his valuable discussions.

\ifCLASSOPTIONcaptionsoff
  \newpage
\fi

\bibliography{bak/bibDB}
\bibliographystyle{IEEEtran}

\begin{IEEEbiographynophoto}{Chuangtao Chen}
	received his B.S. from Shaoguan University in 2019 and his master's degree from Foshan University in 2022. Currently, he is pursuing a Ph.D. degree at Macau University of Science and Technology, specializing in quantum machine learning.
\end{IEEEbiographynophoto}

\begin{IEEEbiographynophoto}{Qinglin Zhao}
	received his Ph.D. degree from the Institute of Computing Technology, the Chinese Academy of Sciences, Beijing, China, in 2005. From May 2005 to August 2009, he worked as a postdoctoral researcher at the Chinese University of Hong Kong and the Hong Kong University of Science and Technology. Since September 2009, he has been with the School of Computer Science and Engineering at Macau University of Science and Technology and now he is a professor. His research interests include blockchain and decentralization computing, machine learning and its applications, Internet of Things, wireless communications and networking, cloud/fog computing, quantum computing and quantum machine learning. He serves as an associate editor of IEEE Transactions on Mobile Computing and IET Communications.
\end{IEEEbiographynophoto}

\begin{IEEEbiographynophoto}{MengChu Zhou}
	(Fellow, IEEE) received his Ph. D. degree from Rensselaer Polytechnic Institute, Troy, NY in 1990 and then joined New Jersey Institute of Technology where he has been Distinguished Professor since 2013. His interests are in Petri nets, automation, robotics, big data, Internet of Things, cloud/edge computing, and AI.  He has 1200+ publications including 17 books, 850+ journal papers (650+ in IEEE transactions), 31 patents and 32 book-chapters. He is Fellow of IFAC, AAAS, CAA and NAI.
\end{IEEEbiographynophoto}

\begin{IEEEbiographynophoto}{Zhimin He}
	received his B.S. and Ph.D. degrees from the School of Computer Science \& Engineering at South China University of Technology, Guangzhou, China, in 2010 and 2015, respectively. He is currently an associate professor in the School of Electronic and Information Engineering at Foshan University. His research interests include quantum machine learning and adversarial learning. He has led projects funded by the National Natural Science Foundation of China and the Guangdong Province Natural Science Foundation. He has published 63 papers in prominent international academic journals and conferences.
\end{IEEEbiographynophoto}

\begin{IEEEbiographynophoto}{Zhili Sun}
	(Senior Member, IEEE) is currently a
	Professor with the 5G\&6G Innovation Centre, University of Surrey, Guildford, U.K. His research interests include wireless and sensor networks, satellite communications, mobile operating systems, traffic engineering, Internet protocols and architecture, QoS, multicast and security. He has published more than 240 papers in international journals and conferences, and three books. He is an Editorial Board Member of Nature-Scientific Report on Space Technology and International Journal of Satellite Communications and Networking.
\end{IEEEbiographynophoto}


\begin{IEEEbiographynophoto}{Dr. Haozhen Situ}
	earned his Ph.D. in Computer Software and Theory from Sun Yat-sen University and is currently an Associate Professor in the Department of Computer Science and Engineering at the School of Mathematics and Informatics (School of Software), South China Agricultural University, where he also serves as a Master's Supervisor. He is an executive member of the Quantum Computing Committee of the China Computer Federation (CCF). Dr. Situ has a long-standing research focus on quantum computing, with 80 SCI papers published. He has also led or participated in several projects funded by the National Natural Science Foundation of China and the Natural Science Foundation of Guangdong Province.
\end{IEEEbiographynophoto}

\clearpage
\includepdf[pages=-,pagecommand={}]{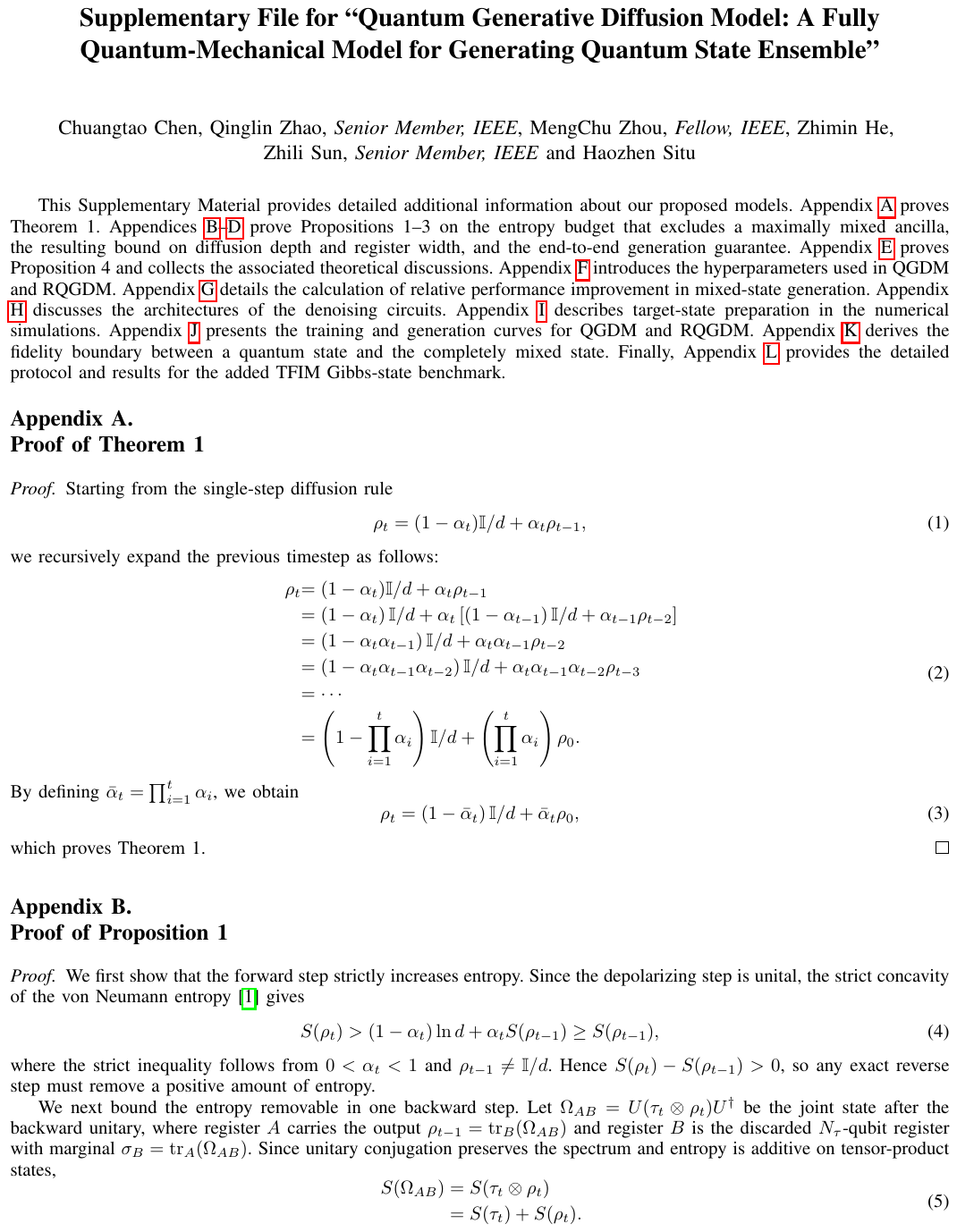}



%
\end{document}